\documentclass[pdflatex,sn-mathphys-num]{sn-jnl}

\usepackage{graphicx}
\usepackage{amsmath,amssymb,amsfonts}
\usepackage{amsthm}
\usepackage{mathrsfs}
\usepackage[title]{appendix}
\usepackage{xcolor}
\usepackage{siunitx} 

\date{March 2026}

\newcommand{\mycomment}[1]{}
\begin{document}

\title[Programmable superconducting diode from nematic domain control in FeSe]{Programmable superconducting diode from nematic domain control in FeSe}

\author*[1]{\fnm{R. D. H.} \sur{Hinlopen}}\email{roemer.hinlopen@mpsd.mpg.de}
\author[1]{\fnm{C.} \sur{Putzke}\dag}
\author[1]{\fnm{L.} \sur{Holeschovsky}\dag}
\author[2]{\fnm{R.} \sur{Nicholls}}
\author[3]{\fnm{F.} \sur{Ronning}}
\author[3]{\fnm{E. D.} \sur{Bauer}}
\author[2,4]{\fnm{N. E.} \sur{Hussey}}
\author[1]{\fnm{P. J. W.} \sur{Moll}}

\affil[1]{Max Planck Institute for Structure and Dynamics of Matter, Hamburg, Germany}
\affil[2]{Wills Physics, University of Bristol, United Kingdom}
\affil[3]{Los Alamos National Laboratory, Los Alamos, NM, USA}
\affil[4]{HFML-FELIX, Nijmegen, Netherlands}
\affil[\dag]{These authors contributed equally}

\abstract{
The superconducting diode effect (SDE) allows polarity-dependent critical currents when time-reversal and current-inverting spatial symmetries are broken. Superconducting diodes show promise for applications, but inversion asymmetry is usually encoded in sample geometry or non-centrosymmetric crystals, rendering them static circuit elements. Here we demonstrate a programmable superconducting diode whose functionality is encoded in correlated electronic domains. We use the nematic superconductor FeSe as a platform and report a large intrinsic SDE with efficiencies up to $\eta \sim 75\%$ due to vortices interacting with nematic twin boundaries. The domain wall configuration thus encodes the SDE of the device. Through intense microsecond current pulses to quench the nematic order at rates exceeding $10^7$~K/s, we modify the domain pattern and control the polarity and strength of the SDE. These results establish a new paradigm in which superconducting circuit elements can be programmed through patterns imprinted into correlated electronic states.
}



\maketitle

The superconducting diode effect (SDE) refers to a polarity dependence of the critical current beyond which supercurrents no longer flow free of dissipation, $I_c^+ \neq |I_c^-|$. Such directional supercurrents allow ideal rectification and its applications form the superconducting analogue of a semiconductor diode. From symmetry considerations the effect is allowed only when time‑reversal symmetry and all spatial symmetries that invert the applied current are broken, such as inversion or mirror operations.

Most realizations achieve this symmetry breaking by design. Artificial heterostructures \cite{Ando2020, Hou2023}, asymmetric Josephson junctions \cite{Wu2022, Golod2022, Diez-Merida2023, Gupta2023} and non‑centrosymmetric crystals \cite{Liu2024} introduce fixed inversion asymmetry that determines the diode polarity. In these systems the rectification direction is directly encoded in the device structure and can typically only be reversed by changing the polarity of an applied magnetic field. 

Unconventional superconductors, however, offer a broader design space when other correlated states coexist with superconductivity. There, the symmetry breaking responsible for the SDE may instead be generated by asymmetries of coexisting orders \cite{Lin2022, Moll2023, Le2024}. If the required asymmetry were created electronically rather than structurally, the diode functionality could become tunable or even programmable.

Here, we report on the discovery of a large intrinsic SDE in electronic-nematic FeSe, link it to electronic nematic domain wall interactions, and demonstrate such superconducting diodes are programmable using microsecond current pulses.

\begin{figure}[h!]
    \centering
    \includegraphics[width=\linewidth]{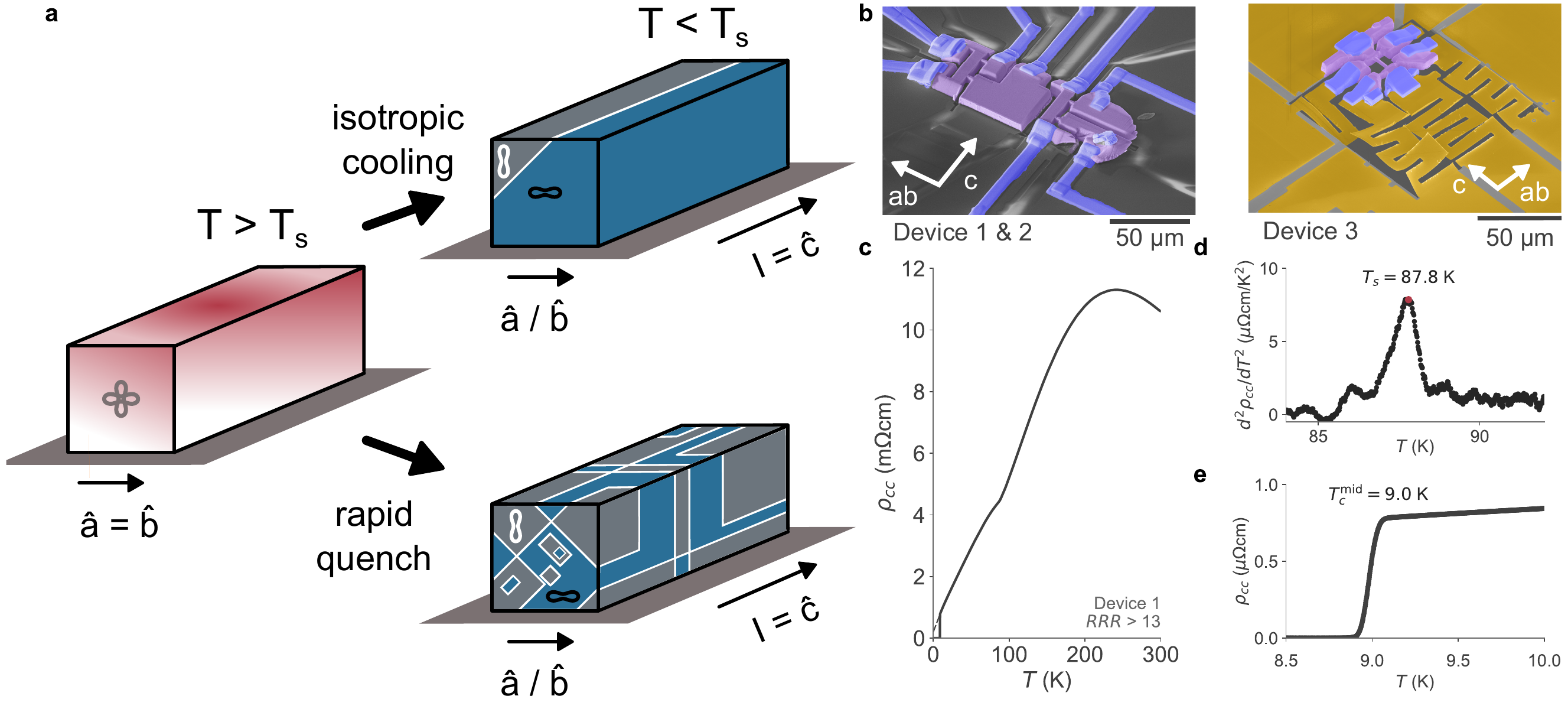}
    \caption{\textbf{Quenching the nematic order in FeSe.} \textbf{a} Adiabatically cooling FeSe microstructures results in a macro-domained electronic nematic state. By contrast, heat quenching FeSe welded to a substrate at 10~MK/s stabilizes a nano-domained state through inhomogeneous thermal expansion. In this fashion, quench rates can be used to control the density and size of the nematic domains. 4-leaf flower and peanut shapes indicate electron pocket shape and orientation \cite{Watson2017, Sprau2017} \textbf{b} Cryogenic FIB-manufactured microstructures used in this study to measure $c$-axis transport. The first two devices are made out of a single lamella, the left labeled device 1, each with cross section $\approx\SI{50}{\um\squared}$. The third device is made from the same single crystal and mounted strain-free with cross section $\approx\SI{5}{\um\squared}$. \textbf{c} Cooldown curve with typical maximum at 250 K, high RRR and strange metal behavior in the $T\rightarrow 0$ limit. \textbf{d} Second derivative showing the nematic transition temperature $T_s$. \textbf{e} Sharp superconducting transition $T_c$ under warming and cooling.}
    \label{fig1}
\end{figure}

FeSe is a layered unconventional superconductor with $T_c=\SI{9}{K}$ that undergoes a nematic transition at $T_s\approx\SI{90}{K}$ without magnetic order \cite{Hsu2008}. Below $T_s$ the electronic structure develops pronounced in‑plane anisotropy \cite{Watson2015prb,Tanatar2016,Watson2017,Bartlett2021}, indicative of the nematic character of the ordered state. The spontaneous symmetry breaking strongly affect the superconducting state, including the 5-fold gap anisotropy \cite{Sprau2017} and elliptical vortex structure \cite{Putilov2019}. The nematic phase produces domains whose domain walls have been shown to strongly suppress superconductivity \cite{Song2012}, which is well understood \cite{Blatter1994} and indeed observed to produce strong vortex pinning \cite{Zhou2023, Terashima2024}.

In order to observe the emergent asymmetric vortex dynamics arising from vortices pinned at nematic domain walls, we require highly symmetric devices to eliminate spatial symmetry breaking through geometric factors and ensure the only source of inversion symmetry breaking arises from the electronic domain walls. For the same reason, we use high quality single crystals to minimize defects. Furthermore, to rapidly quench these devices as well as obtain appreciable current densities, we require micron-sized bars. Finally, the devices must carry current along the $c$-direction such that for in-plane magnetic fields the Lorentz force drives vortices across nematic twin domain boundaries. We fabricate the required devices by cryogenic focused ion beam (FIB) milling (Fig.~\ref{fig1}a,b). 

Cryogenic FIB processing avoids Se loss during fabrication and thereby preserves the stoichiometry and electronic properties of the material. The resulting microstructures reproduce the characteristic transport signatures of bulk FeSe, including the resistivity maximum near \SI{250}{K} (Fig.~\ref{fig1}a), the nematic transition at $T_s=\SI{88}{K}$ (Fig.~\ref{fig1}b) and a sharp superconducting transition at $T_c=\SI{9.0}{K}$ (Fig.~\ref{fig1}c). Residual resistivity ratios comparable to bulk crystals confirm the high quality of the devices (Extended Data~E1) \cite{Kaluarachchi2016}. X-ray photoelectron spectroscopy (XPS, Extended Data~E2) and energy-dispersive X-ray spectroscopy (EDX, Extended Data~E3) were performed to further characterize the structures and show the preservation of Se under cryogenic FIB. All data thus points to a pristine FeSe lattice within the devices, the required starting point to study and control nematic domain formation.

\begin{figure}[h!]
    \centering
    \includegraphics[width=\linewidth]{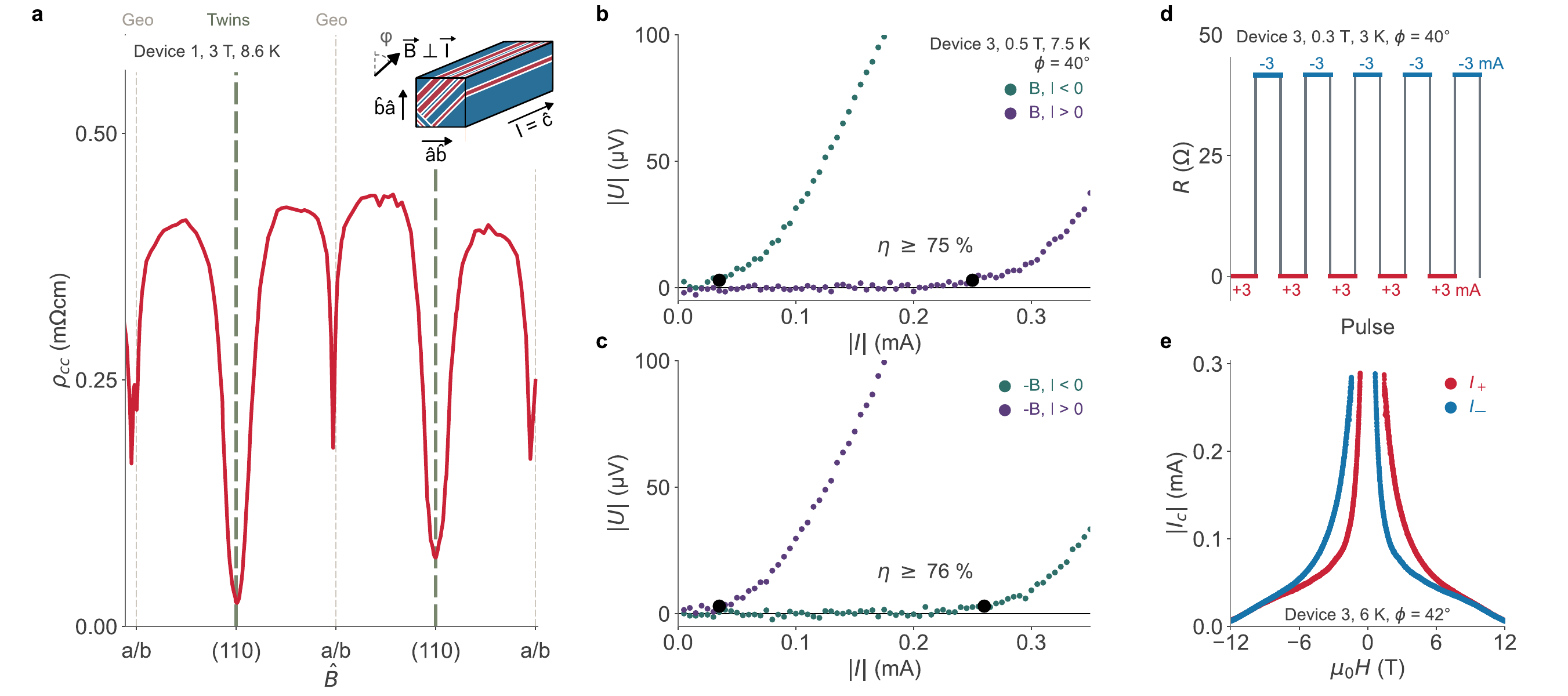}
    \caption{\textbf{Superconducting diode effect at twin pinning directions} \textbf{a} Single-axis in-plane rotation of the magnetic field while in the flux flow regime for the out-of-plane resistivity. Vortex pinning is observed in narrow angle ranges along $a/b$ directions when the magnetic field is along the surface of the structure (marked Geo), while stronger pinning is observed when the magnetic field is near the (110) direction along the nematic twin domain walls. Inset: Magnetic field and current orientation in a macrodomained sample. \textbf{b} At lower temperatures in the superconducting state, a large superconducting diode effect is observed when the magnetic field is near the twin boundary direction. Each datapoint is a separate square current pulse measured within \SI{4}{\us} of the pulse start. The critical current is determined using a fixed voltage marked in black. \textbf{c} Under the same conditions but opposite field polarity, the SDE is inverted. \textbf{d} At even lower temperatures, ideal rectification is observed in a zero-resistance state $<0.04~\Omega$. \textbf{d} The effect decreases with increasing magnetic field, but remains clearly visible at high magnetic fields well beyond the single vortex regime.}
    \label{fig2}
\end{figure}

Angle‑dependent magnetotransport measurements confirm substantial twin boundary vortex pinning. By rotating the field within the $ab$-plane in the flux-flow regime we reveal two alternating pinning directions separated by $45^\circ$ through their characteristic minimum in the flux-flow resistance (Fig.~\ref{fig2}a). Besides the well-known geometric Bean-Livingston barrier when the field is parallel to a flat surface (crystallographic $a$ or $b$ direction), further minima appear for fields parallel to the lattice diagonal. These correspond to vortices aligned with the twin boundaries consistent with previous studies of vortex pinning at nematic domain walls in FeSe \cite{Zhou2023, Terashima2024}. We show magnetic field sweeps of the pinning effect in Extended Data~E4, rotation as a function of temperature in Extended Data~E5 and rotation as a function of magnetic field in Extended Data~E6.

The device is subsequently cooled into the zero-resistance state and current–voltage characteristics are recorded. We use a lock-in method for exploration (Extended Data~E7) and microsecond pulsed IV-curves for detailed data, high currents and minimal heating (Extended Data~E8). When the magnetic field is oriented close to the twin pinning directions, the IV-curves exhibit a pronounced superconducting diode effect (Fig.~\ref{fig2}b,c,d) that is absent in the normal state (Extended Data~E7 and E8). The asymmetry in critical currents is quantified through the diode efficiency

\begin{equation}
\eta=\frac{I_c^+ - |I_c^-|}{I_c^+ + |I_c^-|}
\label{eq:eta}
\end{equation}

All investigated devices display substantial diode efficiencies that depend on temperature, magnetic field and field angle. Under optimized conditions, the SDE reaches efficiencies conservatively estimated at $\eta = \SI{75}{\%}$ (Fig.~\ref{fig2}b,c). Optimization searches were performed on a 2-axis rotator (calibration in Extended Data~E9) for devices 1 and 3, see Extended Data~E10 and E11. We emphasize that the mere appearance of non‑reciprocal transport in such devices is striking because FeSe and the device geometry are structurally centrosymmetric, which renders SDE symmetry-forbidden \cite{Moll2023}. The observation of this large SDE is furthermore restricted to a narrow angle range around the twin boundary where pinning is observed in the flux-flow regime. We thus find direct evidence that the nematic domain boundaries generate the effective inversion symmetry breaking required for the diode effect through domain formation. The domains are tied to the nematic order, which can be manipulated via the comparatively low transition temperature $T_s$. This opens a new design space for superconducting diodes determined by phase-change memory effects. The domain pattern encodes the strength and polarity of the diode at a given field polarity.

A distinctive feature of the SDE in FeSe is the robustness to applied magnetic field. Typically the strongest SDE efficiencies are found at low fields in the \SI{10}{mT} range in the single-vortex regime \cite{Lin2022, Le2024}. In these FeSe devices, the intrinsic diode signal is maximal near 0.5~T and remains observable deep into the collective pinning regime (Fig.~\ref{fig2}e and even above 12~T shown below), reflecting the strong vortex pinning provided by nematic twin boundaries.

Next we turn towards deterministic writing of the SDE via domain patterning. We achieve this by transiently heating the device above the nematic transition using intense microsecond current pulses and rapidly quenching it back to low temperature. The sample itself is used as thermometer to monitor temperature (Extended Data~E12). The quench parameters are used to stabilize distinct metastable nematic domain configurations, which in turn realize distinct asymmetric current-voltage characteristics. While precise control over nematic domain realizations is challenging, the vast space of pulse parameters enables a systematic search for desired diode behaviors (Extended Data~E13). The fast timescale for the switching of the SDE enables functional programmable superconducting diodes. 

\begin{figure}[h!]
    \centering
    \includegraphics[width=\linewidth]{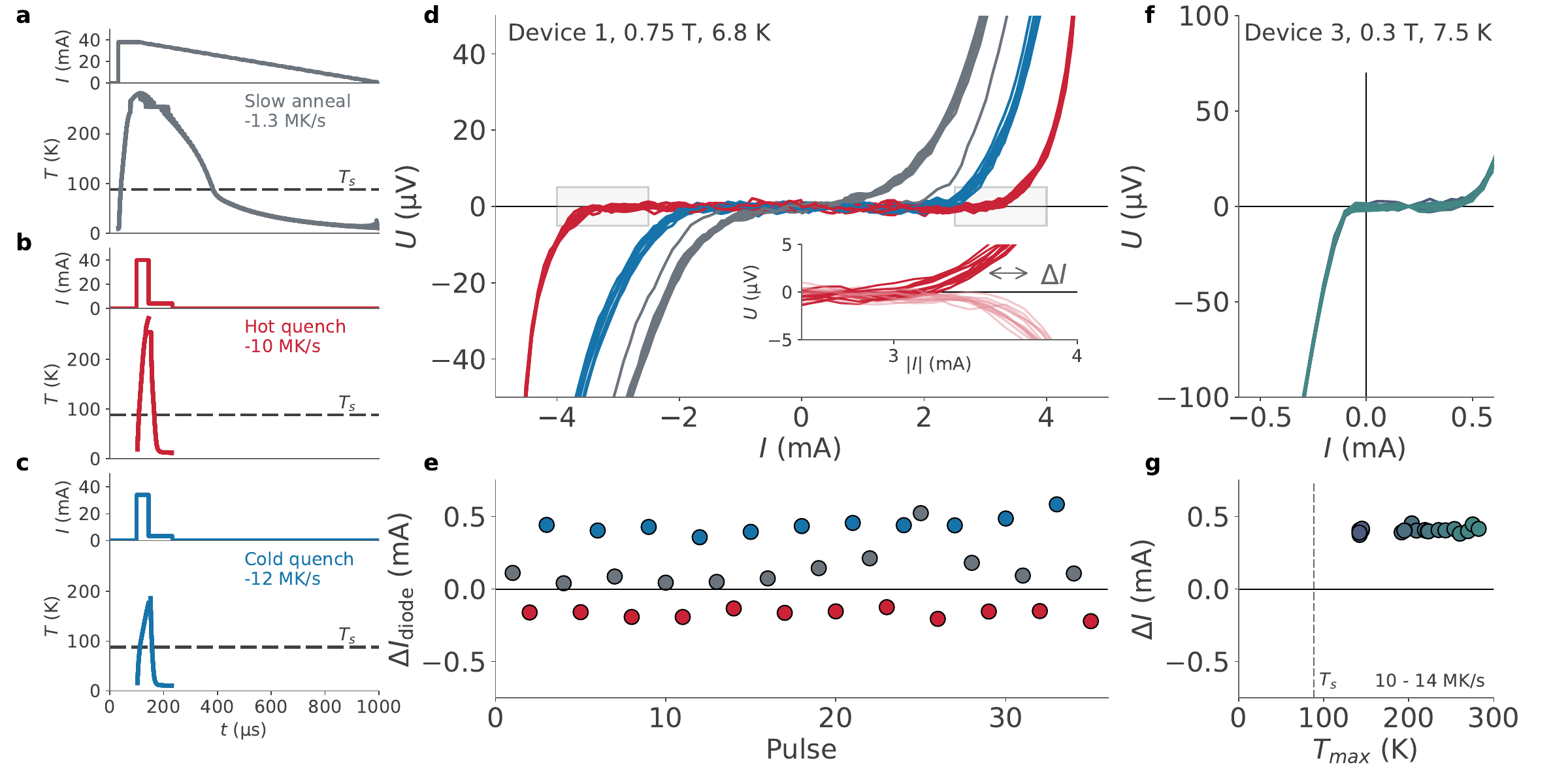}
    \caption{\textbf{Programming the SDE in FeSe}. \textbf{a} Pulse shape for a comparatively slow anneal within 1ms from room temperature, cooling through $T_s$ at \SI{1.3}{MK/s}. The temperature profile is estimated by converting the time-dependent resistance using the cooldown curve. Uncertainty of order \SI{10}{\percent} exists due to thermal inhomogeneity of the sample, most evident near the maximum in $R(T)$ at \SI{250}{K}. \textbf{b} Rapid and hot quench from nearly room temperature at \SI{10}{MK/s}. \textbf{c} Rapid but colder quench from $T_{\mathrm{max}}<\SI{200}{K}$ at similar quench rate. \textbf{d} We cycle through the pulses and alternate each one with square-pulse IV curves to write and read the SDE. The curves are colored based on the directly preceding pulse type. The 35 IV curves form 3 clusters showing deterministic control of the state of the diode effect. Inset: Plotting $|I|$ allows for direct visualization of the inverted SDE after hot quenches. Similar zooms for the other pulses are shown in Extended Data~E14. \textbf{e} The diode effect $I_c^+-|I_c^-|$ extracted from panel \textbf{d}. The hot and cold quench reproducibly stabilize opposite chirality SDEs while the slow anneal results in an approximately diode-less state (except pulse 25). \textbf{f} On the strain-free device we perform quenches with varying peak temperatures alternated with IV curves cooling at \SI{10}{MK/s}. These 23 identical IV curves show this device is unchanged after heat quenching (let alone switching its sign). \textbf{g} The diode effect of the 23 pulses shown in panel \textbf{f} against the peak temperature recorded during their heat pulses.}
    \label{fig3}
\end{figure}

Remarkably, repeated pulse protocols reproducibly reset the device into the same diode state. A three-pulse protocol is used to demonstrate the versatility and robustness of the concept: (I) "Slow annealing" produces a nearly symmetric configuration with minimal SDE response (Fig~\ref{fig3}a). (II) "Hot quenching" from room temperature generates reverse polarity diode behavior compared to a standard cooldown (Fig~\ref{fig3}b). (III) "Cold quenching" cooling at $\SI{12}{MK/s}$ through $T_s$ from an intermediate temperature around \SI{200}{K} yields a forward diode (Fig~\ref{fig3}c). Each pulse therefore writes a specific diode state into the device within microseconds. We cycle through these pulse protocols and perform an IV curve after each pulse to confirm successful writing of the SDE (Fig.~\ref{fig3}d,e). Deterministic switching of the sign and magnitude of the diode response is indeed observed. Importantly, the equilibrium temperature and applied magnetic field to the device remain unchanged. These results demonstrate a microsecond writable superconducting diode in which electronic domain patterns encode programmable circuit states.

\begin{figure}
    \centering
    \includegraphics[width=\linewidth]{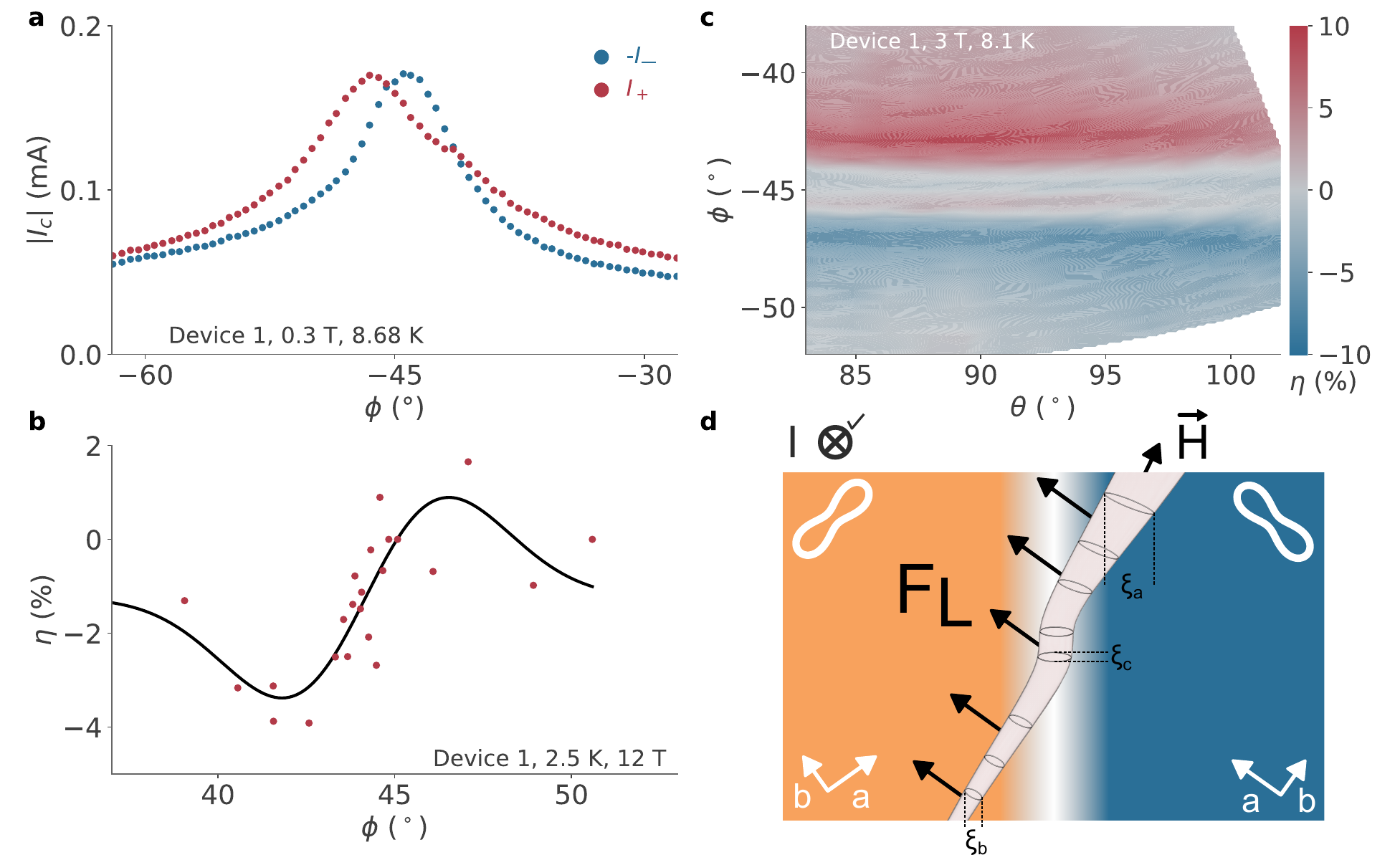}
    \caption{\textbf{Twin boundary origin and leaning vortices}. \textbf{a} Typical angle scan of the SDE showing quasi-Lorentzian critical current enhancement with shift between current polarities. Measured using the lock-in method at low currents. The SDE is inverted when the magnetic field is rotated through the twin domain boundary direction at $\phi=\SI{45}{\degree}$ and diminishes after $\sim \SI{10}{\degree}$ misalignment in $\phi$ direction at \SI{0.3}{T}. \textbf{b} Angle dependence of the SDE using square-pulse IV characteristics at \SI{12}{T} showing the persistence of the diode effect. Black line is a fit to the derivative of a Lorentzian. \textbf{c} Critical current map from 755 IV curves using the lock-in method on a 2-axis rotator. At \SI{3}{T} the twin pinning acceptance angle is about \SI{2}{\degree} in-plane and $>$\SI{10}{\degree} out-of-plane, showing the insensitivity to out-of-plane misalignment. Note the triple sign change in the $\phi$ direction. Calibration of the angles is shown in Extended Data~E9. \textbf{d} Schematic of a leaning vortex with distinct $\xi_a\neq\xi_b$ on either side of the domain wall. The indicated Lorentz force shrinks the vortex and easily depins it compared to the opposite direction (current polarity), causing critical current asymmetry. Peanuts indicate the electron pocket orientation on either side \cite{Watson2017}.}
    \label{fig4}
\end{figure}

Fundamentally, the SDE requires both time‑reversal and spatial symmetry breaking. Time‑reversal symmetry is broken by the magnetic field while spatial symmetry breaking is clearly linked to nematic domain boundaries. However, an idealized straight domain boundary forms a mirror symmetry plane, which forbids an SDE for fields aligned exactly along the domain boundary. A finite angle between the vortices and the domain boundary breaks the mirror plane, as has been demonstrated before in mirror-symmetric devices at oblique field angles \cite{Moll2025}. This important symmetry check is indeed confirmed, as a finite diode response appears only when the magnetic field is slightly misaligned from the domain wall direction (Fig.~\ref{fig4}a,b). By contrast, out-of-plane misalignment barely affects the SDE (Fig.~\ref{fig4}c and Extended Data~E15 and E16).

The critical current asymmetry can be rationalized in terms of elastically deformed vortices at the twin boundaries (Fig~\ref{fig4}d). Any vortex comes with an energy cost due to the suppression of Cooper pair formation in its core. This energy cost per unit length, the vortex line energy $\epsilon$ is proportional to the cross section of a tube of the radius of the coherence length, $\xi$, i.e. $\epsilon \sim \xi^2$. A flux line at an angle with a domain wall naturally extends into both nematic domains, tilting towards $a$ and $b$ directions, respectively. As the vortex threads the nematic electron liquid, however, the coherence length within the FeSe plane is not isotropic anymore, $\xi_a\neq\xi_b$, as evident for example from the ellipticity of vortices detected by tunneling spectroscopy \cite{Putilov2019}. This leads to a left-right difference in line energy across the domain wall, $\epsilon_l \sim \xi_l\xi_c \neq \xi_r\xi_c \sim \epsilon_r$. The resulting depinning asymmetry arising from the elasticity difference provides an unexplored mechanism for strong superconducting diode effects.

Looking ahead, complete control over the diode behavior requires microscopic knowledge and manipulation of the complex nematic domain pattern, which can be addressed through geometry, strain and thermal design within a device. Each of these are expected to provide a symmetry lowering bias that is amplified through kinetic limitation when the sample is quenched through $T_s$. The feasibility of such strain design is experimentally confirmed by low-strain control experiments. Indeed, when the FeSe bar is mounted onto thin SiN$_x$ membranes and decoupled from any thermal mismatch to the substrate (Fig.~\ref{fig1}b), quenching has no measurable effect on the critical current and the device enters the exact same diode state regardless of pulse parameters (Fig.~\ref{fig3}f,g). With improved domain control a vast design space for non-reciprocal superconducting devices emerges.

From a materials perspective, the SDE mechanism found here should not be unique to FeSe. Any anisotropic type‑II superconductor with broken $\mathcal{C}_4$ rotational symmetry must show this SDE when vortices encounter domain walls at an oblique angle to a domain wall. For instance, vortex pinning at twin boundaries is well established in conventional and unconventional superconductors like YBa$_2$Cu$_3$O$_{7-\delta}$ \cite{Blatter1994}. Yet in many cases the asymmetry is driven by crystal structure transitions at elevated temperatures, such as grain boundaries of Nb, which cannot be manipulated in a device. Instead, the symmetry breaking should be associated with a low-temperature electronic order that strongly couples to superconductivity. This ensures simultaneously the switchability by modest current pulses, a substantial suppression of superconductivity at the domain walls and a sizable difference $\xi_a\neq\xi_b$. Naturally, electronic nematicity is prevalent in the larger family of iron-based superconductors such as underdoped BaFe$_2$As$_2$, where nematic and superconducting orders coexist \cite{Bohmer2013}. Other candidate platforms include competing charge-density-wave orders as in kagome superconductors including CsV$_3$Sb$_5$ \cite{Wu2022, Le2024}, spin-density-waves as in organics \cite{Vuletic2002}, magnetically ordered materials as in heavy fermion compounds \cite{Visser1998, Park2006, Khanenko2025} or co-existing superconducting orders such as UTe$_2$ \cite{Wu2026}.


The controllable superconducting diode effect bridges disjunct fields of phase-change memories, superconducting technology and quantum materials. The domain pattern both stores state information and enables amplitude and sign tunability important for analog applications. As all superconducting technology, it is inherently non-linear and controlling this non-linearity is at the heart of this domain control. It will be exciting to see how ideas from high-temperature domain control in memristive and neuromorphic applications based on phase-change, using either charge \cite{Stojchevska2014, Xu2023} or magnetic phases \cite{Kirilyuk2010}, find analogs in superconducting circuits.

\clearpage
\section*{Methods}

Single crystals of FeSe were synthesized via chemical vapor transport. Elemental Fe (\SI{0.201}{g}, \SI{99.998}{\percent} purity) and Se (\SI{0.258}{g}, \SI{99.99}{\percent}) powders were mixed and loaded into a quartz ampule with transport agents KCl (0.501 g, 99.9\%) and AlCl$_3$ (\SI{1.792}{g}, \SI{99.985}{\percent}). The ampule (ID \SI{9}{mm}, length \SI{15}{cm}) was evacuated, sealed and heated to \SI{415}{\degreeCelsius} (source zone) and \SI{345}{\degreeCelsius} (growth zone) in a multi-zone furnace. After \SI{30}{days}, the ampule was cooled to room temperature and platelet-like single crystals were collected from the cold end of the ampule.

For device fabrication we used a Thermofisher Ga FIB for devices~1 and 2 and Thermofisher Helios 5 Xe plasma FIB for device~3. All devices were made using a Kleindiek MHCS cold stage insert at \SI{-60}{\degreeCelsius} for both lamella preparation and microstructuring except deposition, which was done at room temperature but well away from the measured section of the device. Only one device (Extended Data~E1) had a degraded RRR of 7, $T_s=\SI{85.5}{K}$ and $T_c=\SI{8.6}{K}$, either due to reactivity while glue mounting or due to heating while Ar etching and high power Au sputtering, both of which were avoided on all other devices by directly contacting using FIB-induced Pt deposition. A Rigaku X-ray diffraction machine was used to orient the crystal and verify single crystallinity. EDX in the Ga FIB and XPS using the Thermofisher NEXSA G2 were used to characterize the FIB damage to the FeSe and highlighted the need for cryogenic FIB to obtain high quality devices, see Extended Data~E2 and E3. 

For the flux flow measurements, cooldowns and 2-axis rotator heatmaps we used a standard 4-point lock-in resistivity measurement with low excitation current $<\SI{100}{\uA}$ to avoid heating. We used a Synktex MCL1-540 lock-in with ability to record IV characteristics at \SI{1}{MHz} frequency while applying a 5-\SI{30}{Hz} AC excitation to the sample (Extended Data~E7). 

For pulsed-IV measurements and microsecond quenching we used a Keithley 6221 AC current source triggered by a Keithley DMM7510 digital multimeter. We obtain 1~MHz full-differential voltage readout and a typical rise time of the sample response of 2-\SI{3}{\us} (Extended Data~E8). We apply 20-\SI{50}{\us} pulses but in data analysis use the readout after \SI{4}{\us} to minimize heating. The only exception is Fig.~\ref{fig2}c, which was evaluated after \SI{30}{\us} at the end of the applied pulses. 

Single-axis rotation experiments were performed using an in-house built rotator. 2-axis rotation was achieved using an Attocube atto3DR with ANC300 controller. Magnetic field was applied using a dry \SI{12}{T} Oxford Instruments Teslatron for devices 1 and 2, and a \SI{16}{T} Oxford Instruments wet system for device 3.

\clearpage 
\bibliography{references}

@article{Ando2020,
   author = {F. Ando and Y. Miyasaka and T. Li and J. Ishizuka and T. Arakawa and Y. Shiota and T. Moriyama and Y. Yanase and T. Ono},
   pages = {373},
   journal = {Nature},
   title = {{Observation of superconducting diode effect}},
   volume = {584},
   year = {2020},
}

@article{Gupta2023,
   author = {M. Gupta and G. V. Graziano and M. Pendharkar and J. T. Dong and C. P. Dempsey and C. Palmstrøm and V. S. Pribiag},
   pages = {3078},
   journal = {Nat. Comm.},
   title = {{Gate-tunable superconducting diode effect in a three-terminal Josephson device}},
   volume = {14},
   year = {2023},
}

@article{Wu2022,
   author = {H. Wu and Y. Wang and Y. Xu and P. K. Sivakumar and C. Pasco and U. Filippozzi and S. S. P. Parkin and Y.-J. Zeng and T. McQueen and M. N. Ali },
   pages = {653},
   journal = {Nature},
   title = {{The field-free Josephson diode in a van der Waals heterostructure}},
   volume = {604},
   year = {2022}
}

@article{Hsu2008,
   author = {F.-C. Hsu and J.-Y. Luo and K.-W. Yeh and T.-K. Chen and T.-W. Huang and P. M. Wu and Y.-C. Lee and Y.-L. Huang and Y.-Y. Chu and D.-C. Yan and M.-K. Wu},
   pages = {14262},
   journal = {PNAS},
   title = {{Superconductivity in the PbO-type structure $\mathrm{\alpha}$-FeSe}},
   volume = {105},
   year = {2008}
}

@article{Zhou2023,
   author = {N. Zhou and Y. Sun and Q. Hou and T. Sakakibara and X. Z. Xing and C. Q. Xu and C. Y. Xi and Z. S. Wang and Y. F. Zhang and Y. Q. Pan and B. Chen and X. Luo and Y. P. Sun and X. Xu and T. Tamegai and M. Xu and Z. Shi},
   pages = {101195},
   journal = {Mat. Tod. Phys.},
   title = {{Intrinsic pinning of FeSe$_{\mathrm{1-x}}$S$_{\mathrm{x}}$ single crystals probed by 
torque magnetometry}},
   volume = {37},
   year = {2023}
}

@article{Song2012,
   author = {C.-L. Song and Y.-L. Wang and Y.-P. Jiang and L. Wang and K. He and X. Chen and J. E. Hoffman and X.-C. Ma and Q.-K. Xue},
   pages = {137004 },
   journal = {Phys. Rev. Lett.},
   title = {{Suppression of superconductivity by twin boundaries in FeSe}},
   volume = {109},
   year = {2012}
}

@article{Terashima2024,
   author = {T. Terashima and H. Fujii and Y. Matsushita and Shinya Uji and Y. Matsuda and T. Shibauchi and S. Kasahara},
   pages = {014518 },
   journal = {Phys. Rev. B},
   title = {{Transport evidence for twin-boundary pinning of superconducting vortices in FeSe}},
   volume = {109},
   year = {2024}
}

@article{Blatter1994,
   author = {G. Blatter and M. V. Feigel'man and V. B. Geshkenbein and A. I. Larkin and V. M. Vinokur},
   pages = {1125 },
   journal = {Rev. Mod. Phys},
   title = {{Vortices in high-temperature superconductors}},
   volume = {66},
   year = {1994}
}

@article{Moll2023,
    author = {P. J. W. Moll and V. B. Geshkenbein},
    title = {{Evolution of superconducting diodes}},
    journal = {Nat. Phys},
    year = 2023,
    volume = 19,
    pages = 1379
}

@article{Bartlett2021,
	author = {J. M. Bartlett A. Steppke and S. Hosoi and H. Noad and J. Park and C. Timm and T. Shibauchi and A. P. Mackenzie and C. W. Hicks},
	title = {{Relationship between transport anisotropy and nematicity in FeSe}},
	journal = {Phys. Rev. X},
	year = {2021},
	volume = {11},
	pages = {021038},
}

@article{Sprau2017,
	author = {P. O. Sprau and A. Kostin and A. Kreisel and A. E. B\"ohmer and V. Taufour and P. C. Canfield and S. Mukherjee and P. J. Hirschfeld and B. M. Anderson and J. C. S\'eamus Davis},
	title = {{Discovery of orbital-selective Cooper pairing in FeSe}},
	journal = {Science},
	year = {2017},
	volume = {357},
	pages = {75},
}

@article{Bohmer2013,
   author = {A. E. B\"{o}hmer and F. Hardy and F. Eilers and D. Ernst and P. Adelmann and P. Schweiss and T. Wolf and C. Meingast},
   pages = {180505},
   journal = {Phys. Rev. B},
   title = {{Lack of coupling between superconductivity and orthorhombic distortion in stoichiometric single-crystalline FeSe}},
   volume = {87},
   year = {2013}
}

@article{Putilov2019,
   author = {A. V. Putilov and C. Di Giorgio and V. L. Vadimov and D. J. Trainer and E. M. Lechner and J. L. Curtis and M. Abdel-Hafiez and O. S. Volkova and A. N. Vasiliev and D. A. Chareev and G. Karapetrov and A. E. Koshelev and A. Y. Aladyshkin and A. S. Mel’nikov and M. Iavarone},
   pages = {144514},
   journal = {Phys. Rev. B},
   title = {{Vortex-core properties and vortex-lattice transformation in FeSe}},
   volume = {99},
   year = {2019}
}

@article{Kaluarachchi2016,
	author = {U. S. Kaluarachchi and V. Taufour and A. E. B\"ohmer and M. A. Tanatar and S. L. Bud'ko and V. G. Kogan and R. Prozorov and P. C. Canfield},
	journal = {Phys. Rev. B},
	pages = {064503},
	title = {{Nonmonotonic pressure evolution of the upper critical field in superconducting FeSe}},
	volume = {93},
	year = {2016},
}

@article{Watson2015prb,
	author = {M. D. Watson and T. K. Kim and A. A. Haghighirad and N. R. Davies and A. McCollam and A. Narayanan and S. F. Blake and Y. L. Chen and S. Ghannadzadeh and A. J. Schofield and M. Hoesch and C. Meingast and T. Wolf and A. I. Coldea},
	title = {{Emergence of the nematic electronic state in FeSe}},
	journal = {Phys. Rev. B},
	year = {2015},
	volume = {91},
	pages = {155106},
}

@article{Watson2017,
	author = {M. D. Watson and A. A. Haghighirad and L. C. Rhodes and M. Hoesch and T. K. Kim},
	title = {{Electronic anisotropies revealed by detwinned angle-resolved photoemission spectroscopy measurements of FeSe}},
	journal = {New J. Phys.},
	year = {2017},
	volume = {19},
	pages = {103021},
}

@article{Tanatar2016,
   author = {M. A. Tanatar and A. E. B\"{o}hmer and E. I. Timmons and M. Schütt and G. Drachuck and V. Taufour and K. Kothapalli and A. Kreyssig and S. L. Bud’ko and P. C. Canfield and R. M. Fernandes and R. Prozorov},
   pages = {127001},
   journal = {Phys. Rev. Lett.},
   title = {{Origin of the resistivity anisotropy in the nematic phase of FeSe}},
   volume = {117},
   year = {2016}
}

@article{Liu2024,
	author = {F. Liu and Y. M. Itashashi and S. Aoki and Y. Dong and Z. Wang and N. Ogawa and T. Ideue and Y. Iwasa},
	pages = {eado1502},
	journal = {Sci. Adv.},
	title = {{Superconducting diode effect under time-reversal symmetry}},
	volume = {10},
	year = {2024}
}

@article{Hou2023,
	author = {Y. Hou and F. Nichele and H. Chi and A. Lodesani and Y. Wu and M, F. Ritter and D, Z. Haxell and M. Davydova and S. Ili$\check{c}$ and O. Glezakou-Elbert and A. Varambally and F. S. Bergeret and A. Kamra and L. Fu and P. A. Lee and J. S. Moodera},
	pages = {027001},
	journal = {Phys. Rev. Lett.},
	title = {{Ubiquitous superconducting diode effect in superconductor thin films}},
	volume = {131},
	year = {2023}
}

@article{Golod2022,
	title = {{Demonstration of a superconducting diode-with-memory, operational at zero magnetic field with switchable nonreciprocity}},
	year = 2022,
	author = {T. Golod and V. M. Krasnov},
	journal = {Nat. Comm.},
	pages = {3658},
	volume = 13
}

@article{Diez-Merida2023,
	author = {J. D\'iez-M\'erida and A. D\'iez-Carl\'on and S. Y. Yang and Y.-M. Xie and X.-J.Gao and J. Senior and K. Watanabe and T. Taniguchi and X. Lu and A. P. Higginbotham and K. T. Law and D. K. Efetov},
	pages = {2396},
	journal = {Nat. Comm.},
	title = {{Symmetry-broken Josephson junctions and superconducting diodes in magic-angle twisted bilayer graphene}},
	volume = {14},
	year = {2023}
}

@article{Lin2022,
   author = {J. X. Lin and P. Siriviboon and H. D. Scammell and S. Liu and D. Rhodes and K. Watanabe and T. Taniguchi and J. Hone and M. S. Scheurer and J. I. A. Li},
   pages = {1221},
   journal = {Nat. Phys.},
   title = {{Zero-field superconducting diode effect in small-twist-angle trilayer graphene}},
   volume = {18},
   year = {2022}
}

@article{Le2024,
   author = {T. Le and Z. Pan and Z. Xu and J. Liu and J. Wang and Z. Lou and X. Yang and Z. Wang and Y. Yao and C. Wu and X. Lin},
   pages = {64},
   journal = {Nature},
   title = {{Superconducting diode effect and interference patterns in kagome CsV$_3$Sb$_5$}},
   volume = {630},
   year = {2024}
}

@article{Wu2026,
   author = {Z. Wu and H. Chen and M. Long and D. Shaffer and D. V. Chichinadze and A. Cabala and T. I. Weinberger and A. J. Hickey and J. Pu and D. Graf and V. Sechovsk\'y and M. Vali$\mathrm{\check{s}}$ka and G. Li and R. Zhou and F. M. Grosche and A. G. Eaton},
   pages = {},
   journal = {{arXiv:2603.02450}},
   title = {{Discovery of an electrically-controllable superconducting memory effect}},
   volume = {},
   year = {2026}
}

@article{Khanenko2025,
	author = {P. Khanenko and J. F. Landaeta and S. Ruet and T. Lühmann and K. Semeniuk and M. Pelly and A. W. Rost and G. Chajewski and D. Kaczorowski and C. Geibel and S. Khim and E. Hassinger and M. Brando},
	pages = {L060501},
	journal = {Physical Review B},
	title = {{Phase diagram of CeRh$_2$As$_2$ for out-of-plane magnetic field}},
	volume = {112},
	year = {2025}
}

@article{Visser1998,
	author = {A. de Visser and R.J. Keizer and M.J. Graf and A.A. Menovsky and J.J.M. Franse},
	pages = {287},
	journal = {J. Mag. Mag. Mat.},
	title = {{On the interplay of small-moment magnetism and superconductivity in UPt$_3$}},
	volume = {177},
	year = {1998}
}

@article{Vuletic2002,
	author = {T. Vuleti\'c and P. Auban-Senzier and C. Pasquier and S. Tomi\'c, D. J\'erome and M. H\'eritier and K. Bechgaard},
	pages = {319},
	journal = {Eur. Phys. J. B},
	title = {{Coexistence of superconductivity and spin density wave orderings in the organic superconductor (TMTSF)$_2$PF$_6$}},
	volume = {25},
	year = {2002}
}

@article{Park2006,
	author = {T. Park and F. Ronning and H. Q. Yuan and M. B. Salamon and R. Movshovich and J. L. Sarrao and J. D. Thompson},
	pages = {65},
	journal = {Nature},
	title = {{Hidden magnetism and quantum criticality in the heavy fermion superconductor CeRhIn$_5$}},
	volume = {440},
	year = {2006}
}

@article{Moll2025,
	author = {P. J. W. Moll},
	pages = {73},
	journal = {Comm. Mat.},
	title = {{Geometrical design of 3D superconducting diodes}},
	volume = {6},
	year = {2025}
}

@article{Kirilyuk2010,
	author = {A. Kirilyuk and A. V. Kimel and T. Rasing},
	pages = {2731},
	journal = {Review of Modern Physics},
	title = {{Ultrafast optical manipulation of magnetic order}},
	volume = {82},
	year = {2010}
}

@article{Stojchevska2014,
	author = {L. Stojchevska and I. Vaskivskyi and T. Mertelj and P. Kusar and D. Svetin and S. Brazovskii and D. Mihailovic},
	pages = {177},
	journal = {Science},
	title = {{Ultrafast switching to a stable hidden quantum state in an electronic crystal}},
	volume = {344},
	year = {2014}
}

@article{Xu2023,
	author = {C. Xu and C. Jin and Z. Chen and Q. Lu and Y. Cheng and B. Zhang and F. Qi and J. Chen and X. Yin and G. Wang and D. Xiang and D. Qian},
	pages = {1265},
	journal = {Nat. Comm.},
	title = {{Transient dynamics of the phase transition in VO$_2$ revealed by mega-electron-volt ultrafast electron diffraction}},
	volume = {14},
	year = {2023}
}

\clearpage
\pagestyle{empty}
\setcounter{figure}{0}
\renewcommand{\thefigure}{E\arabic{figure}}
\renewcommand{\figurename}{Extended Data}

\begin{figure}
    \centering
    \includegraphics[width=\linewidth]{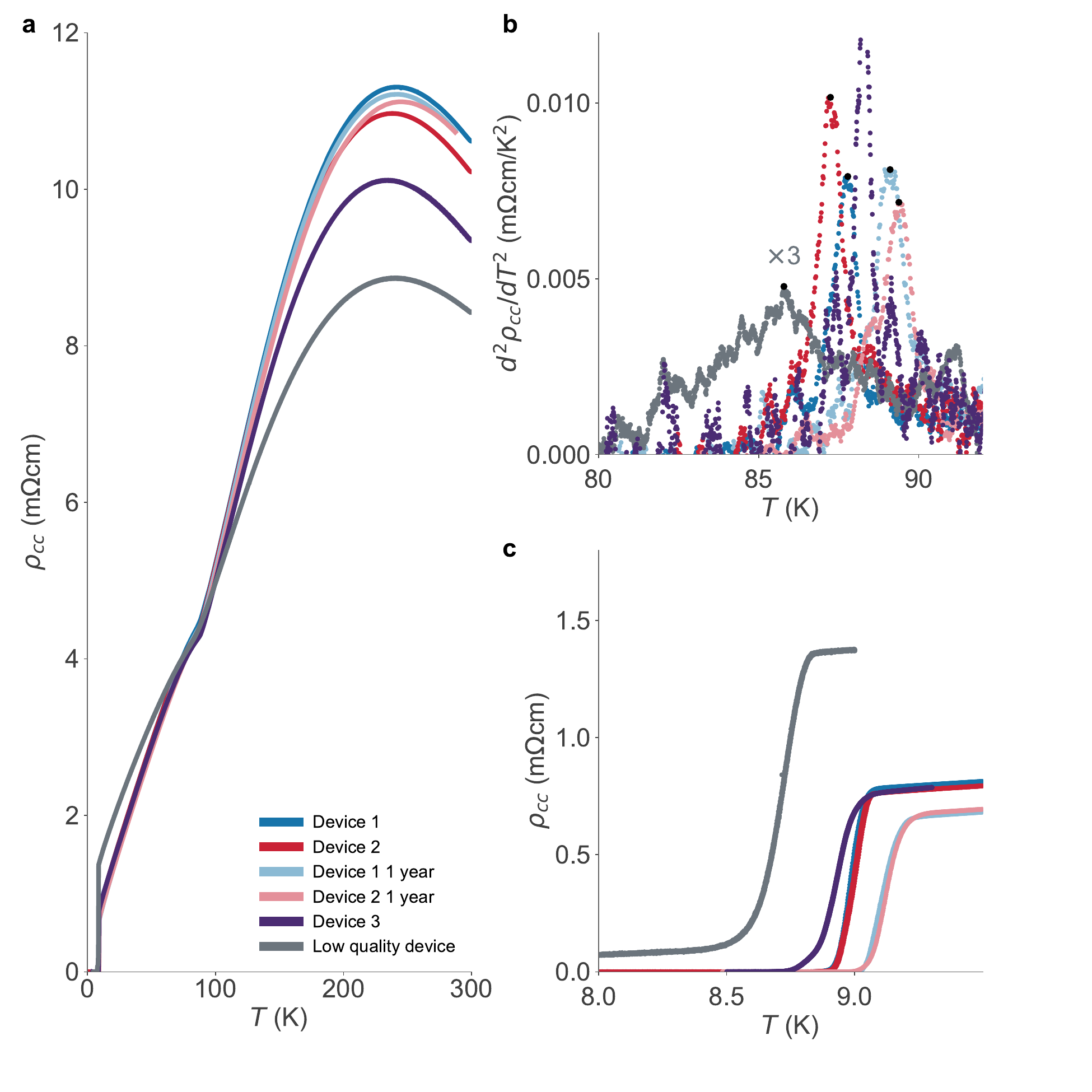}
    \caption{\textbf{Cooldown curves}. \textbf{a} The cooldown curves are shown for device~1 and 2 both directly after fabrication as well as a year onward. The latter not only tests for aging of the microstructures, but is also after all ballpark 1000 current pulses to room temperature. We observe a slight increase in both $T_s$ and $T_c$ over this period, but no broadening of either feature. Device~3 (strain free) shows a similar RRR, $T_s$ and $T_c$. The gold-coated and glue-mounted sample shows a reduced RRR, wide nematic transition and lowest $T_s$ and was therefore not measured further. \textbf{b} Second derivative of the resistivity showing the nematic transition temperature $T_s$. The low-quality device has a broadened transition. \textbf{c} Superconducting transitions and zero resistance for all devices used in this study. }
\end{figure}

\clearpage
\begin{figure}
    \centering
    \includegraphics[width=\linewidth]{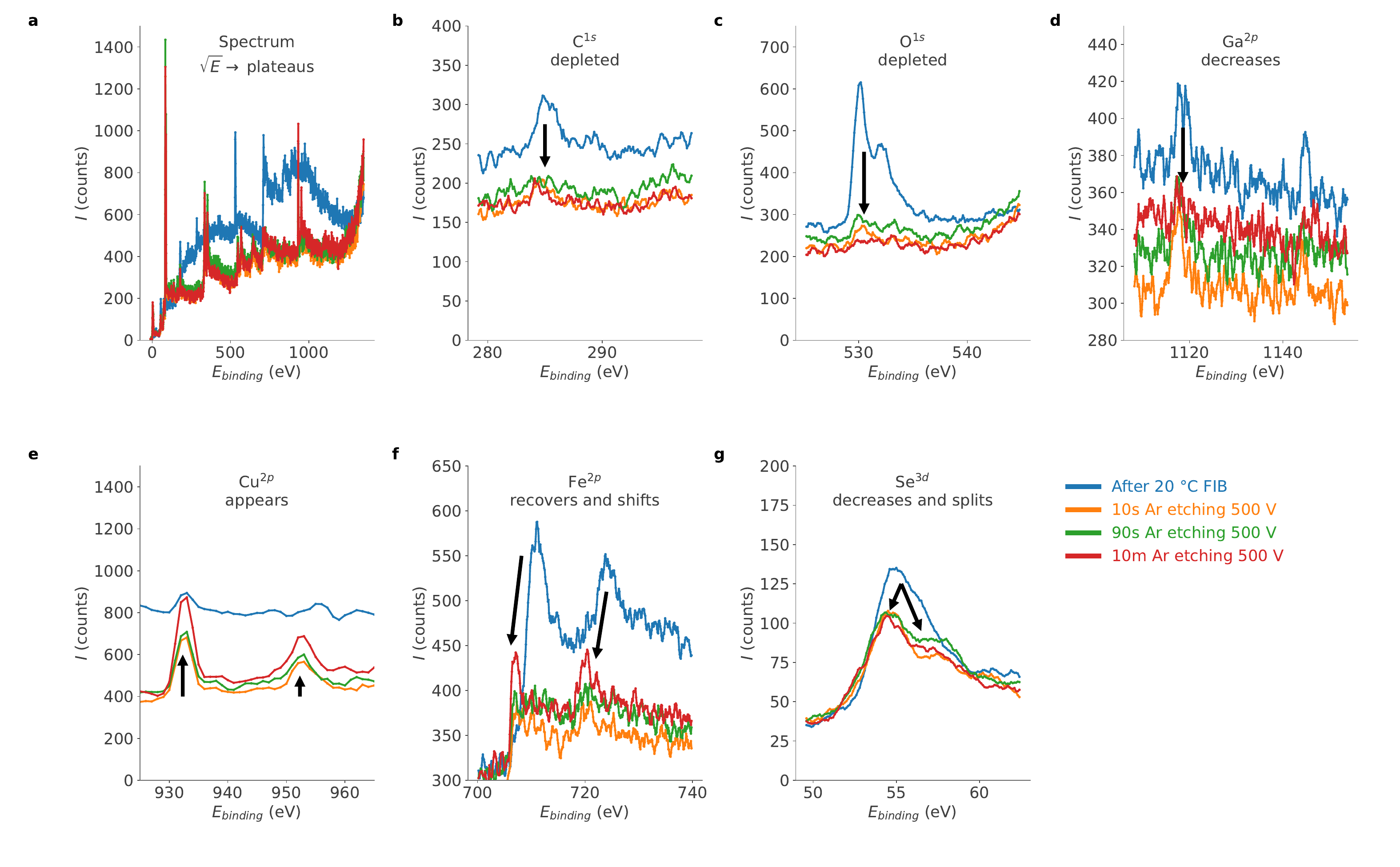}
    \caption{\textbf{FIB-induced loss of Se in XPS}. We use X-ray spectroscopy at \SI{1486.68}{eV} to characterize the surface of FIB-polished samples at \SI{30}{kV} and ambient temperature. \textbf{a} Full spectra alternated with low-voltage Ar etching as indicated in the legend. \textbf{b} The primary carbon peak due to electron microscopy depletes rapidly under Ar etching at \SI{500}{kV}. We estimate \SI{10}{nm/min} of material is removed. \textbf{c} Similarly, oxygen is only observed at the surface layer and the microstructures are indeed stable over a year in a desiccator. \textbf{d} A weak gallium implantation signal is observable and diminishes over about 100~nm depth. Any associated doping is considered minimal given the resistivity measurements match literature results. \textbf{e} Copper contamination is absent before etching and rises afterwards due to Cu redeposition from the transmission electron microscopy (TEM) grid used to mount the sample. \textbf{f} The iron peak decreases due to Cu redeposition. The Fe peak shifts when the surface layer is removed from values which are close to Fe$_2$O$_3$ towards pure Fe, matching the removal of oxygen. \textbf{g} The selenium peak decreases similarly as the Fe peak, which we understand from the Cu redeposition. The peak also sharpens sufficiently to show the two underlying 3d orbitals, indicating a more homogeneous electronic environment of the Se in the bulk compared to the amorphous surface. Data taken on a lamella from the same parent crystal which was used to make devices~1-3.}
\end{figure}

\clearpage
\begin{figure}
    \centering
    \includegraphics[width=\linewidth]{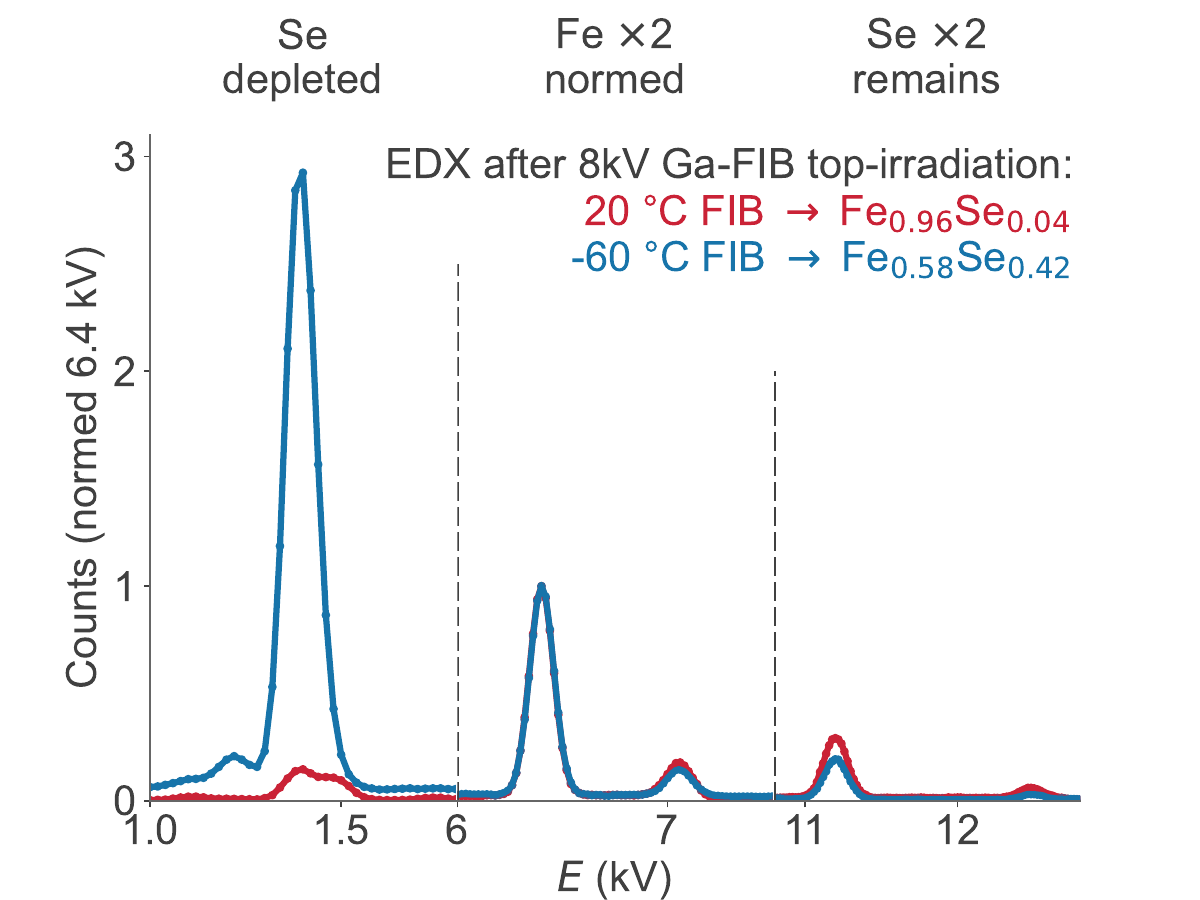}
    \caption{\textbf{FIB-induced loss of Se in EDX}. We use an Oxford Instruments Xplore~30 to measure the EDX signal in the FIB. Here we show two such spectra on the same crystal from which devices~1-3 were fabricated. The red spectrum is obtained after top-irradiation with a Ga FIB at ambient temperature. The blue spectrum on the same area of the same crystal but at \SI{-60}{\degreeCelsius}. The electron energy is \SI{20}{kV} in both cases. The indicated stoichiometry is automatically determined by the AZtec software provided and shows clear preservation of Se under cryogenic conditions. On closer inspection, we observe a near total depletion in the \SI{1.5}{kV} Se signal, while the \SI{11}{kV} Se peak is mildly reduced relative to the Fe peak at \SI{6.4}{kV}. We understand this dichotomy by the difference in escape depth at the indicated energies: The \SI{1.5}{kV} peak is surface sensitive and hence shows primarily the amorphization layer deplete of Se under ambient temperature FIB while at \SI{11}{kV} bulk Se is also measured. This result shows the importance of cryogenic FIB to preserve stoichiometry.}
\end{figure}

\clearpage
\begin{figure}
    \centering
    \includegraphics[width=\linewidth]{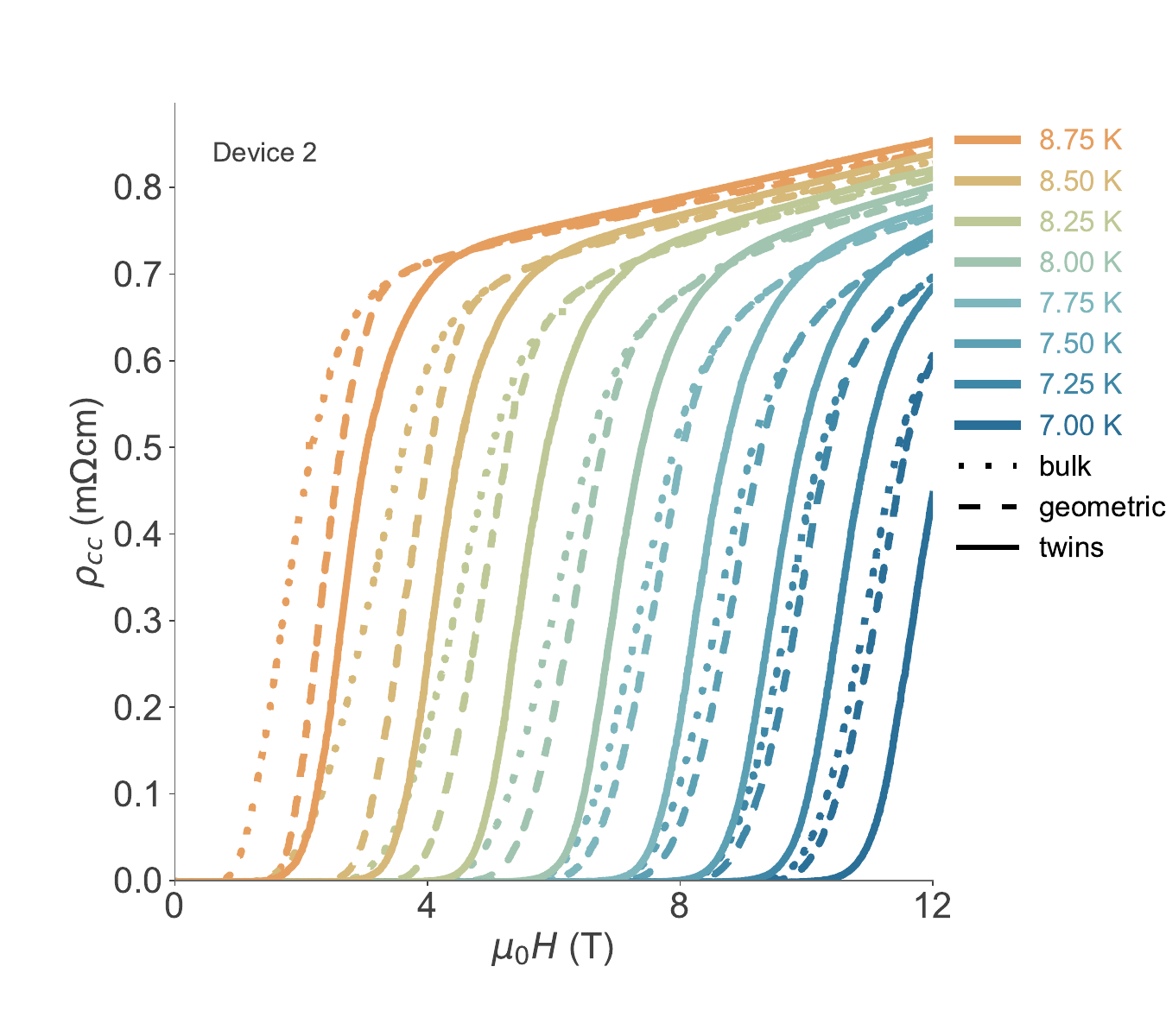}
    \caption{\textbf{Magnetic field sweeps showing vortex pinning} Fieldsweeps at various temperatures near $T_c$ reveal a clear dependence on magnetic field orientation. When the magnetic field is aligned with the surface of the structure, the depinning field is increased relative to an arbitrary in-plane magnetic field direction (marked bulk), while near \SI{12}{T} the bulk and surface depinning field is similar showing reduced effect of the surface. By contrast, when the magnetic field is aligned with the twin domain walls the depinning field is highest at all magnetic temperatures within our field range. The nematic domain wall pinning is both stronger at all applied magnetic fields and more robust to collective pinning than surface effects.}
\end{figure}

\clearpage
\begin{figure}
    \centering
    \includegraphics[width=\linewidth]{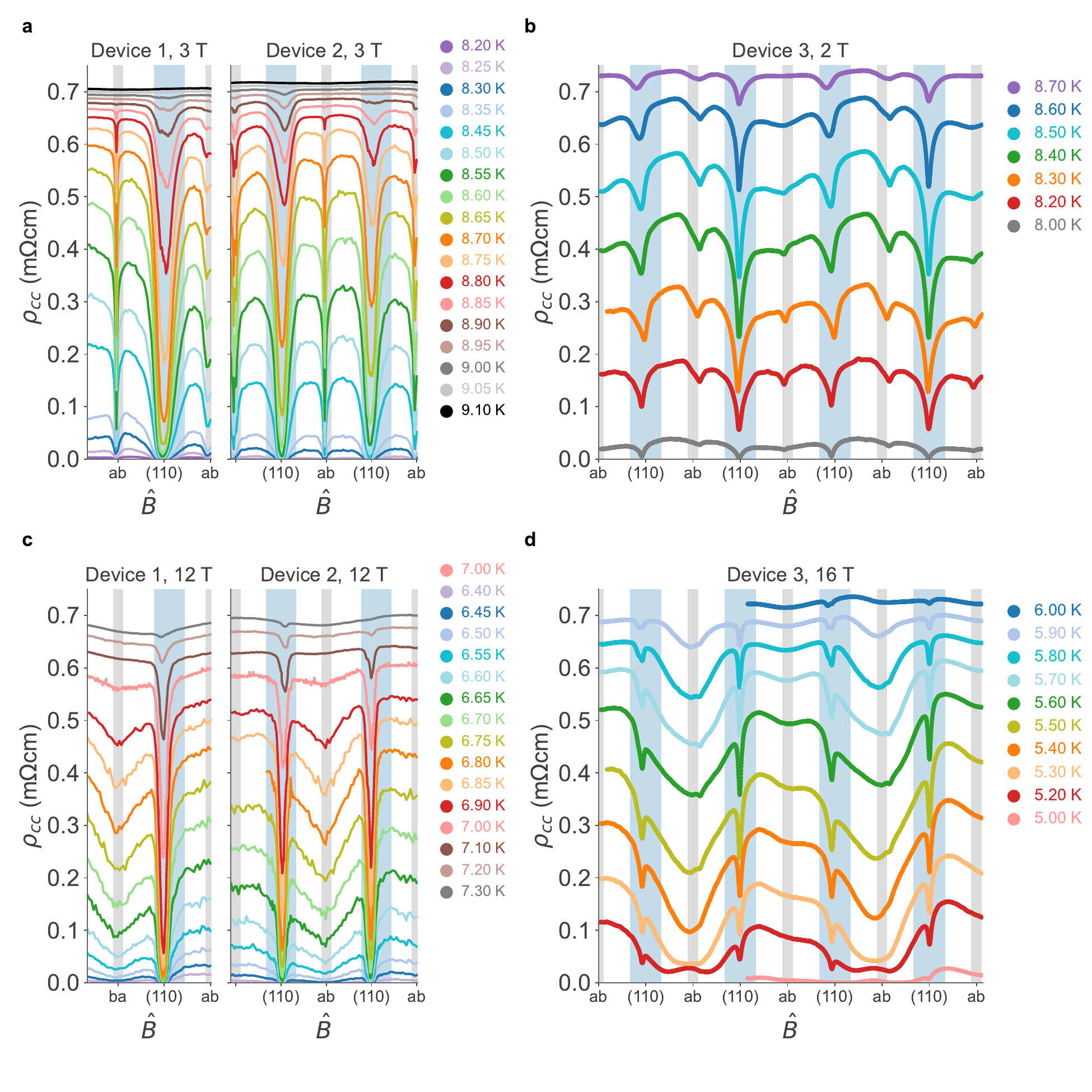}
    \caption{\textbf{Temperature dependence of vortex pinning}. \textbf{ab} The magnetic field is rotated while in the flux flow regime. The resistivity reveals vortex pinning in all three devices. The sharp vortex pinning in $a/b$ direction are due to the surface of the devices, the (110) pinning sites are along the nematic domain walls. As a function of temperature the curves obtains natural vertical offsets due to the recovery of the normal metallic resistance in the flux flow regime without variation in width or strength. \textbf{cd} Same but at maximum field applied. We observe no noticeably surface pinning deep into the collective regime. The gradual background may be due to thermal variation as the thermometer is on the platform of this single-axis rotator. Angles are determined using a Hall probe next to the sample. }
\end{figure}

\clearpage
\begin{figure}
    \centering
    \includegraphics[width=\linewidth]{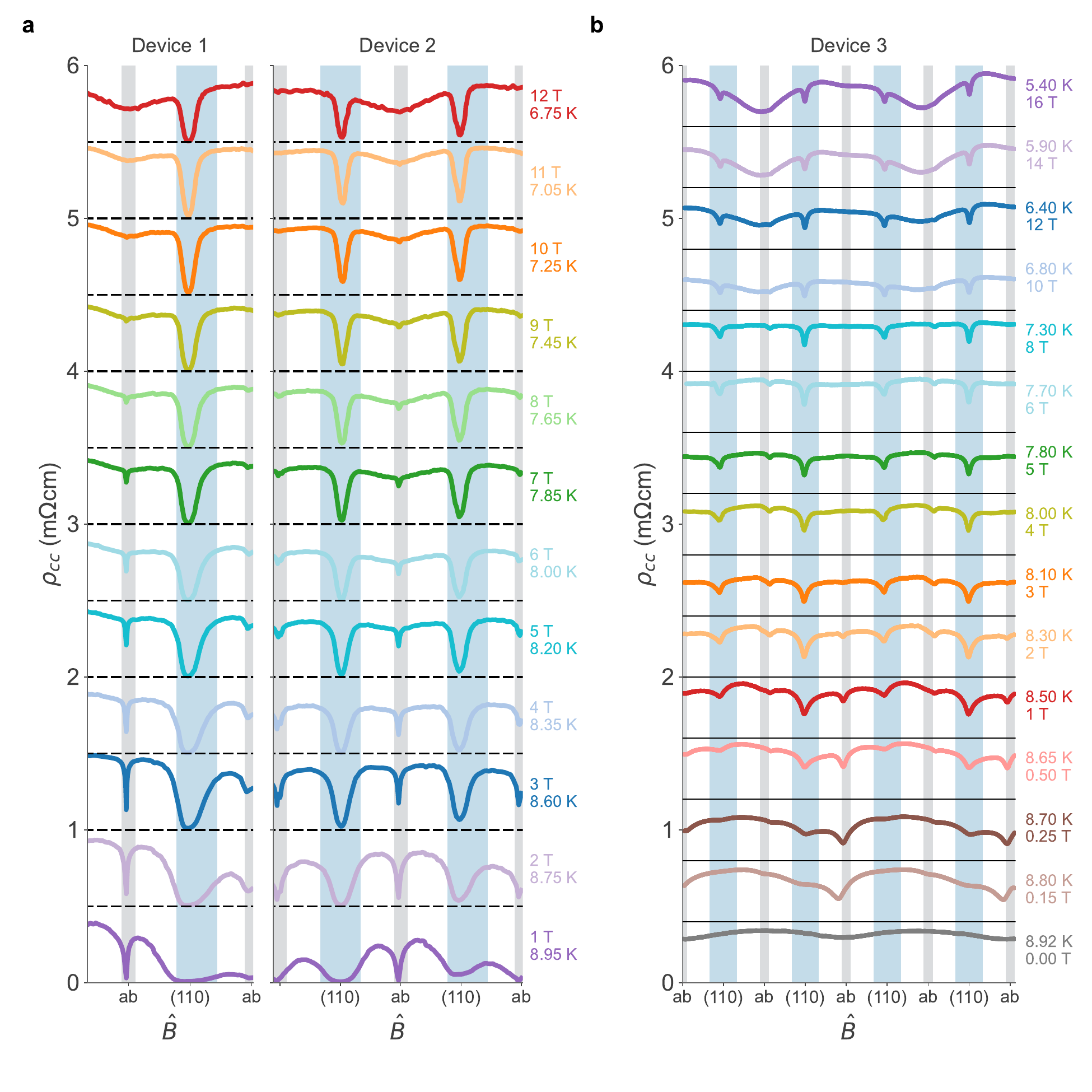}
    \caption{\textbf{Magnetic field dependence of vortex pinning}. The magnetic field is rotated in-plane while measuring $\rho_{cc}$. The temperature at each magnetic field is chosen such that the resistance is approximately half of the normal state value. We observe sharp peaks where the applied magnetic field is aligned with the surface of the structure (geometric, the $a$ and $b$ directions). These peaks decay with increasing field. By contrast, when the magnetic field is aligned with the twin boundary direction $(110)$ we observe stronger and wider pinning which persists to maximum applied magnetic field in all samples. device~3 was later cut into a diamond shape to further verify the sharp peaks originate from the surface. These measurements were taken on a single-axis rotator with thermometer on the platform, which results in a slight temperature variation as evidenced by the zero-Tesla curve due to differences in thermal radiation and exchange gas flow. All other measurements were performed on a 2-axis rotator with thermometer off the platform to avoid this issue and no conclusions are affected.}
\end{figure}

\clearpage
\begin{figure}
    \centering
    \includegraphics[width=\linewidth]{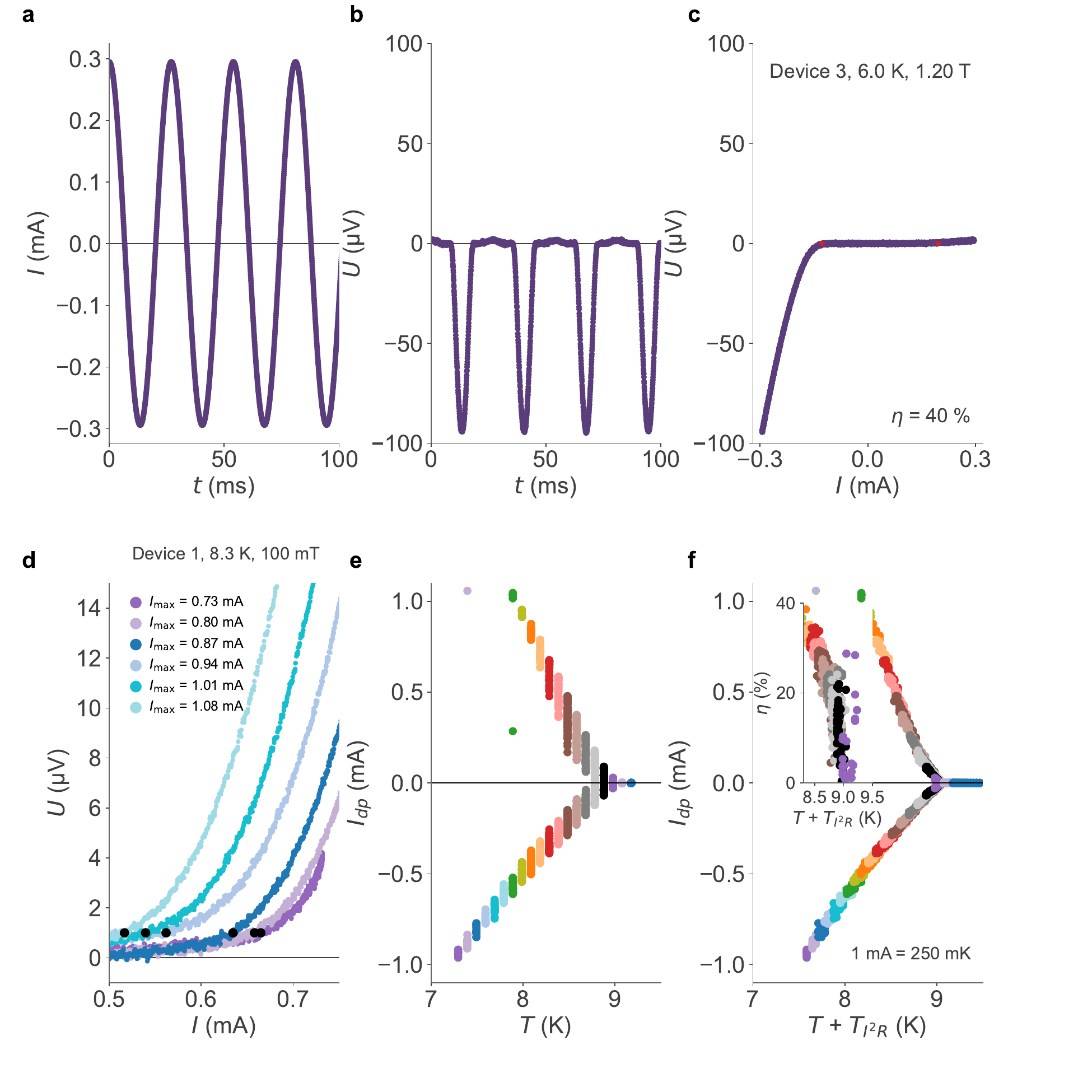}
    \caption{\textbf{Lock-in method for IV curves and heating}. \textbf{a} Example applied current to the device~3. \textbf{b} The resulting voltage shows rectification. \textbf{c} The data plotted as IV curve. \textbf{d} Under identical conditions, higher peak currents result in lower critical currents, showing the persistence of heating throughout the lock-in cycle. \textbf{e} The critical current without applied magnetic field using various maximum currents in the lock-in method. \textbf{f} Using $I^2R$ as a measure for the heat we observe all curves from panel \textbf{e} collapse using the estimate \SI{250}{mK} heating at \SI{1}{mA}. The data is well described by neglecting instantaneous heating and focus on the peak current applied during a cycle.}
\end{figure}

\clearpage
\begin{figure}
    \centering
    \includegraphics[width=\linewidth]{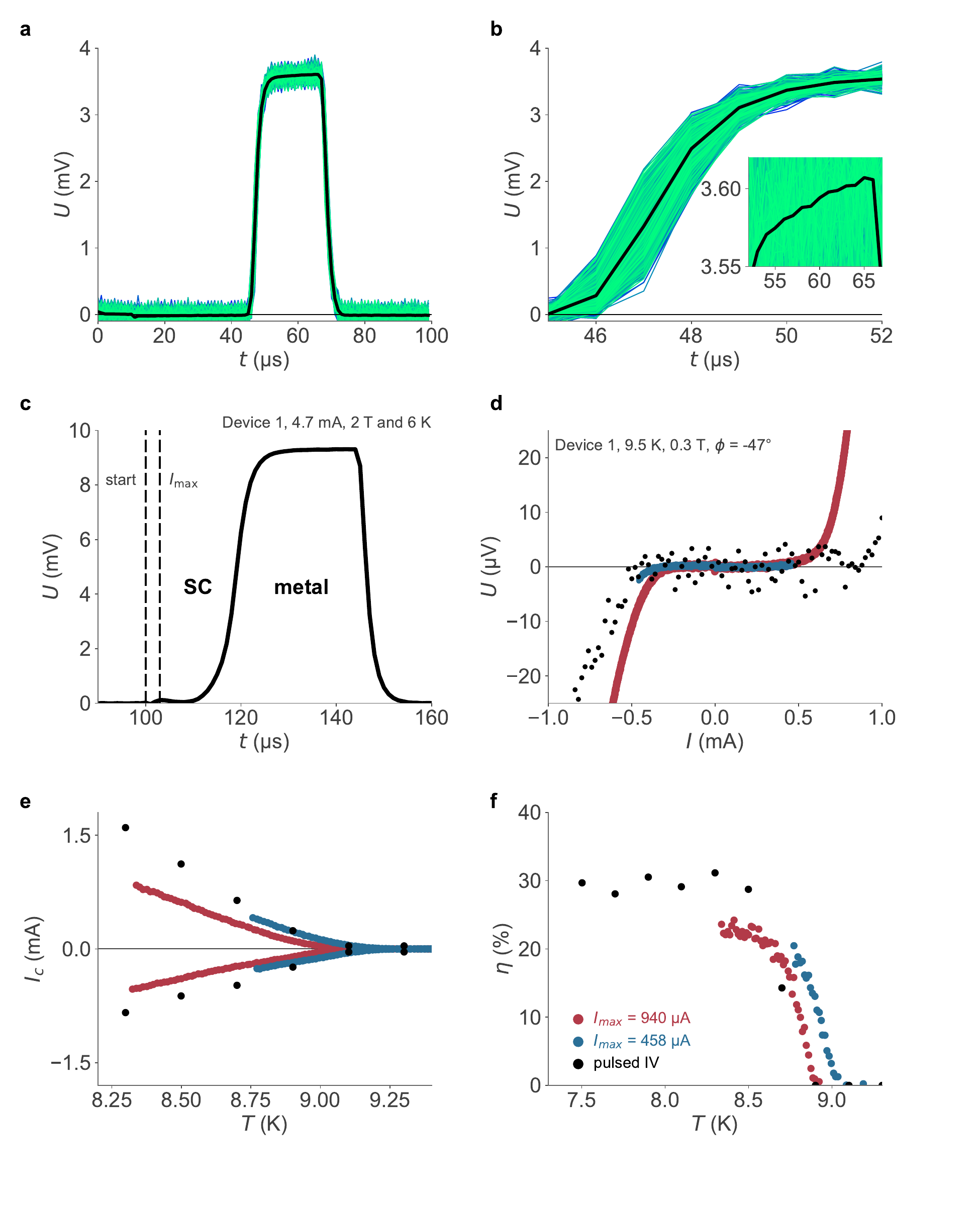}
    \caption{\textbf{Current pulse characteristics and heating reduction}. \textbf{a} Example pulse made with Keithley 6221 AC current source and measured with DMM 7510 Keithley multimeter. We obtain a fully differential voltage readout every $\mu$s. The figure shows 800 pulses in shades of green-blue which are averaged at each timestamp to obtain the black curve. \textbf{b} Zooms on the same data. The main panel shows a rise time of 2-\SI{3}{\us} and the inset accuracy down to \SI{1}{\uV} when using many repeats. Note that the finite slope over time is due to heating. \textbf{c} Reduction in heating. The sample is cooled into the superconducting state at \SI{6}{K} with $T_c$ at \SI{8.7}{K} under \SI{2}{T} applied magnetic field. A large current pulse is applied which heats the sample through the contacts before reaching the depinning current after $~\sim$\SI{10}{\us}. Subsequently, the device heats up and turns entirely metallic due to thermal runaway. Using the lock-in method, a current of \SI{4.7}{mA} is estimated to heat the sample to \SI{11.5}{K}. By reading the voltage \SI{4}{\us} after the start of the pulse, we thus reduce the impact of heating using the pulsed-IV set-up. \textbf{d} IV curves taken under the same conditions using the lock-in method (blue and red with different maximum currents) and square-pulse IV (in black). The critical current is higher using pulsed IV due to reduced heating. \textbf{e} The data in panel \textbf{d} is taken as a function of temperature to show the reduced heating from the pulsed IV set-up. \textbf{f} The diode efficiency for the data in panel \textbf{e}. Due to additional heating at higher critical currents, the diode efficiency is underestimated by the lock-in method. Note that the pulsed IV data is taken over a wide range of current up to \SI{4}{mA}, resulting in low resolution very close to $T_c$.}
\end{figure}

\clearpage
\begin{figure}
    \centering
    \includegraphics[width=\linewidth]{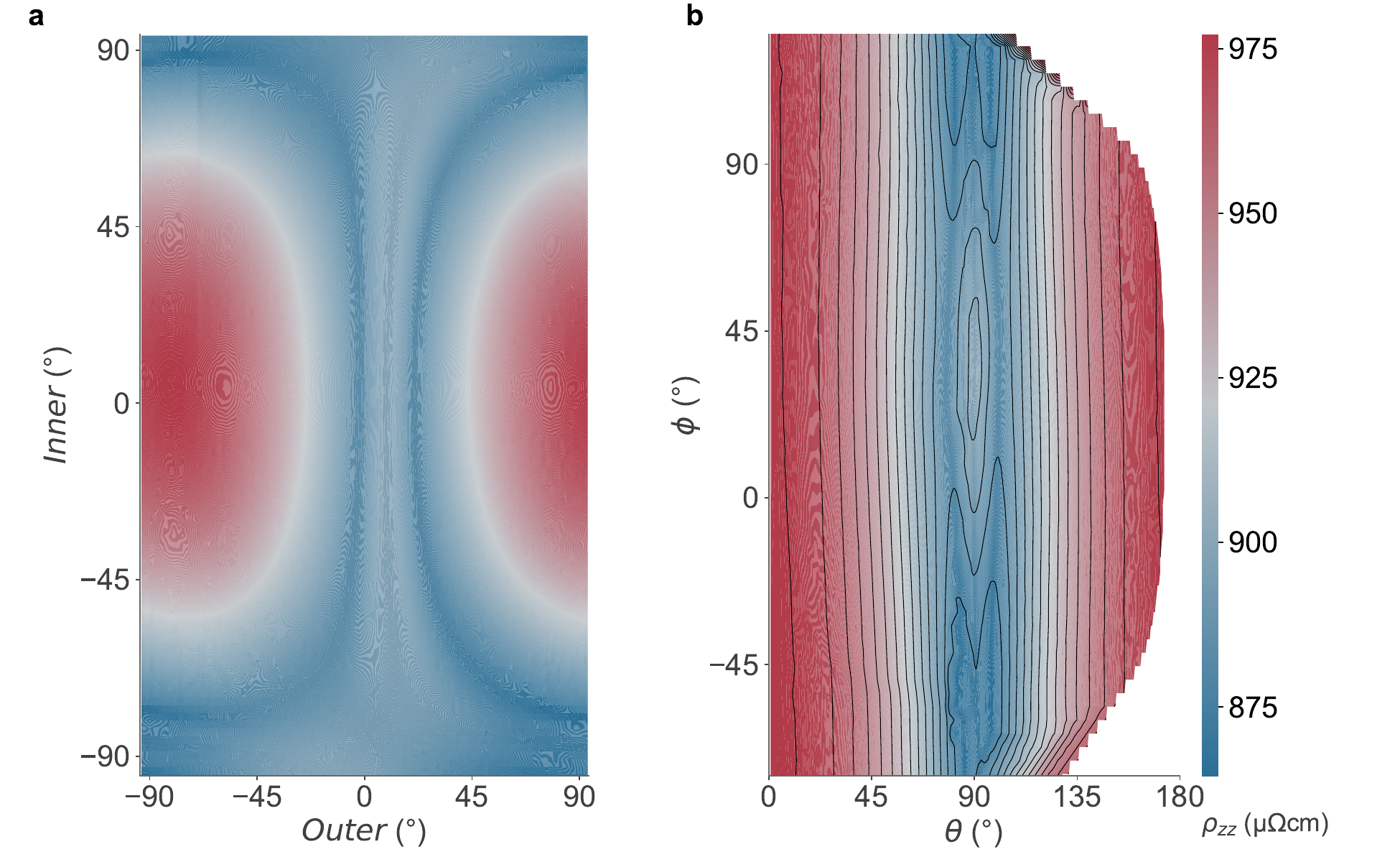}
    \caption{\textbf{Angle calibration of 2-axis rotator}. On the Attocube 2-axis rotator we measure as a function of an inner and outer angle, which are measured using slip-stick motion. Here we show the angle dependent magnetoresistance (ADMR) for device~1 at \SI{10}{K} and \SI{12}{T}. The resistance is maximal for $H//I//c$. Meanwhile, a local maximum is observed for $H//ab$, known as the coherence peak and evidence of the cleanliness of the device. To obtain the angles in the crystal frame ($\theta$, $\phi$), we apply a \ang{6} offset to the outer angle, followed by rotation around the $z$-axis, then $y$-axis and finally again the $z$-axis. These 4 degrees of freedom are fit such that the coherence peak appears at $\theta=\ang{90}$ for all azimuthal angles. Finally, $\phi$ is shifted such that the twin domain wall pinning is at $\pm$\ang{45}. Data shown is for device~1.}
\end{figure}

\clearpage
\begin{figure}
    \centering
    \includegraphics[width=\linewidth]{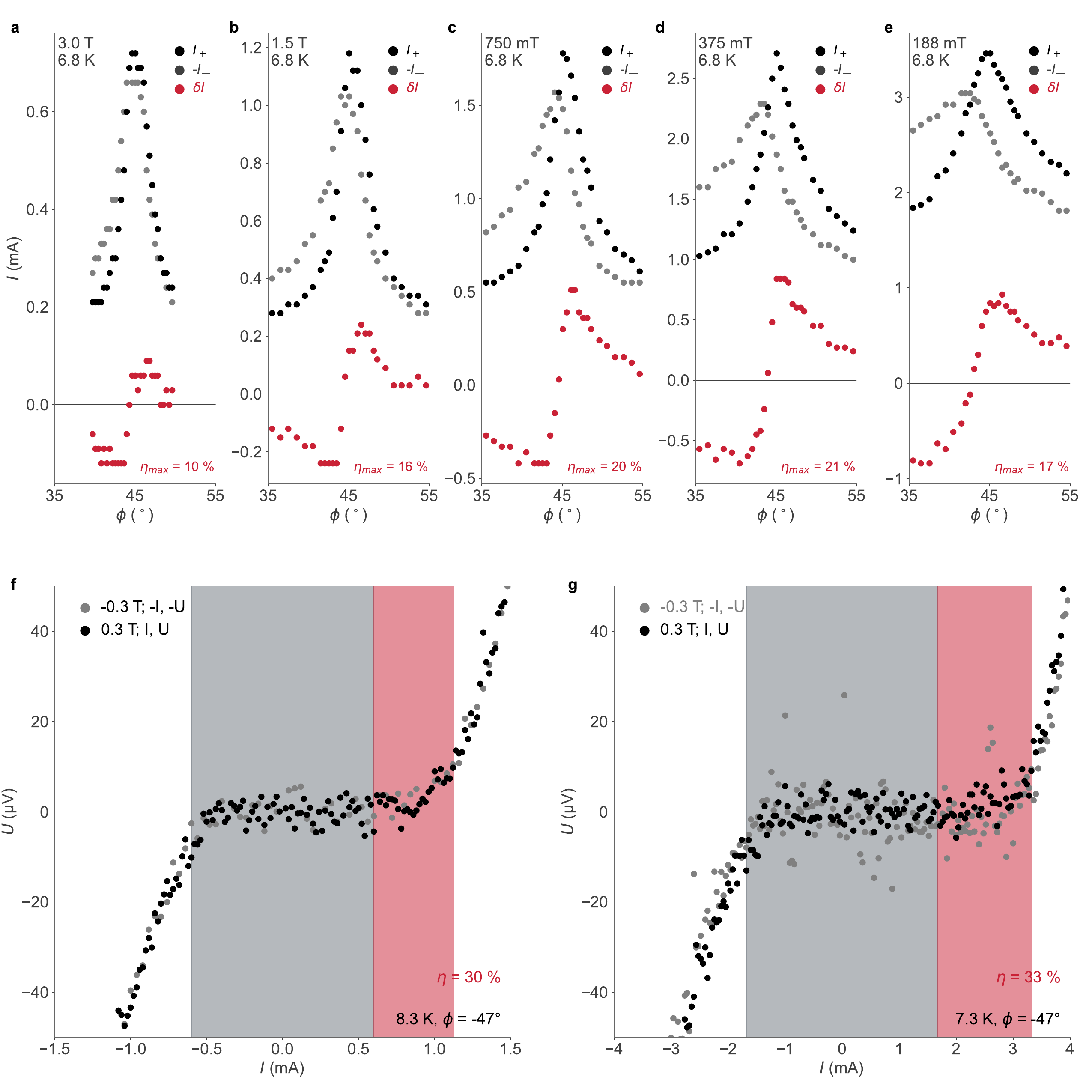}
    \caption{\textbf{Optimizing diode efficiency in device~1}. \textbf{a}-\textbf{e} Angle dependence of the critical current rotating through the twin domain wall using the pulsed-IV 2-axis rotator setup to maintain $\theta=$\ang{90}. The peak efficiency is largest at low applied magnetic fields, but saturates below \SI{750}{mT}. The sample reset by warming up the cryostat before these measurements. \textbf{fg} Detailed IV curves with the maximum efficiency SDE obtained in device~1 of about \SI{33}{\percent}, which happened to be slightly stronger at the opposite twin boundary than the survey in panels \textbf{a}-\textbf{e}. }
\end{figure}

\clearpage
\begin{figure}
    \centering
    \includegraphics[width=\linewidth]{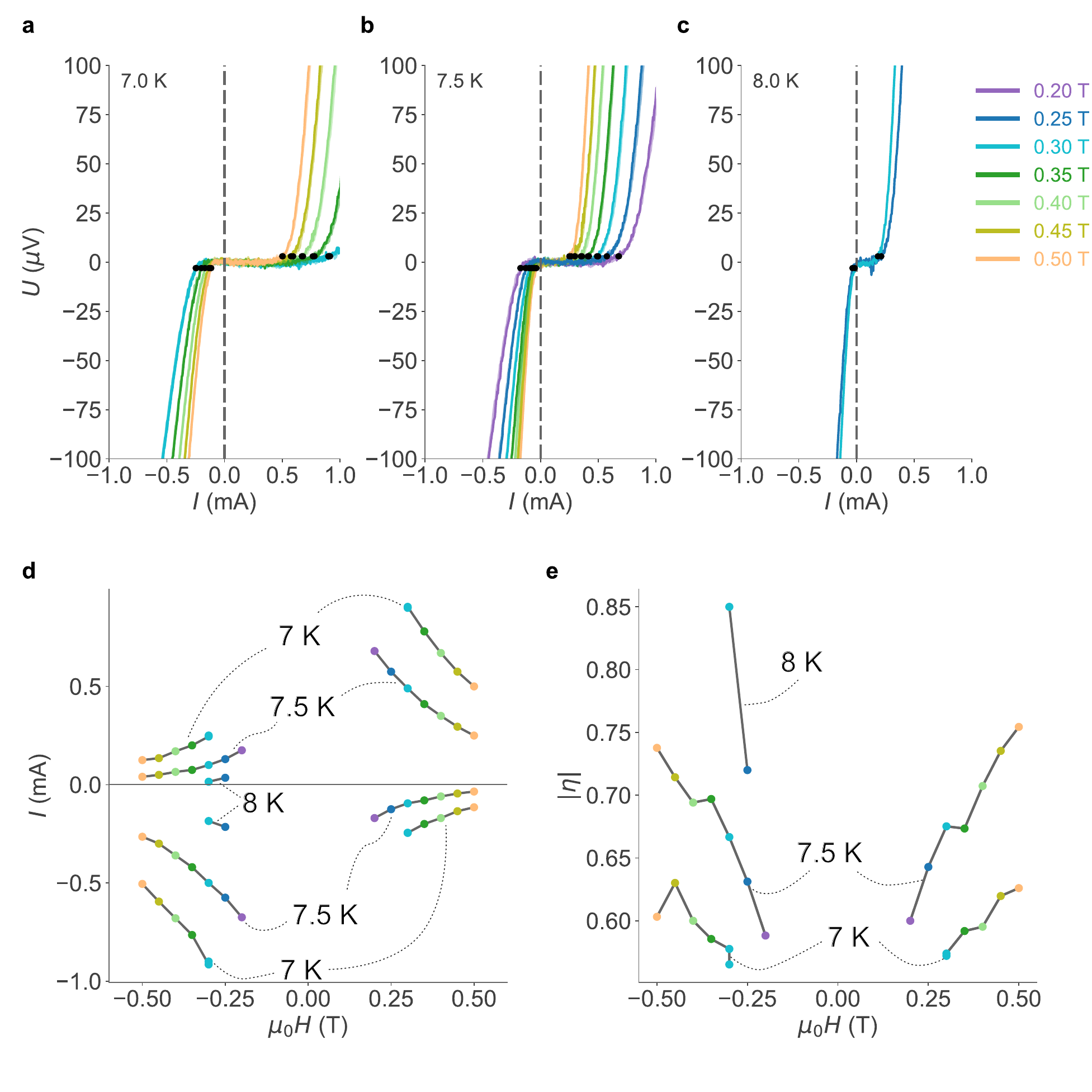}
    \caption{\textbf{Optimizing diode efficiency in device~3}. Panels \textbf{a}, \textbf{b}, \textbf{c} show IV curves obtained at 3 different temperatures as a function of magnetic field showing strong diode effects. Negative polarity magnetic fields are shown with $-I$ and $-U$ and show odd symmetry as expected. Black marked points indicate the critical current using a constant-voltage criterion. \textbf{d} Overview of the critical current as a function of applied magnetic field and temperature. \textbf{e} Close to $T_c$ the efficiency slowly increases in device~3. Data at \SI{0.5}{T} and \SI{7.5}{K} is shown in the main text.}
\end{figure}

\clearpage
\begin{figure}
    \centering
    \includegraphics[width=\linewidth]{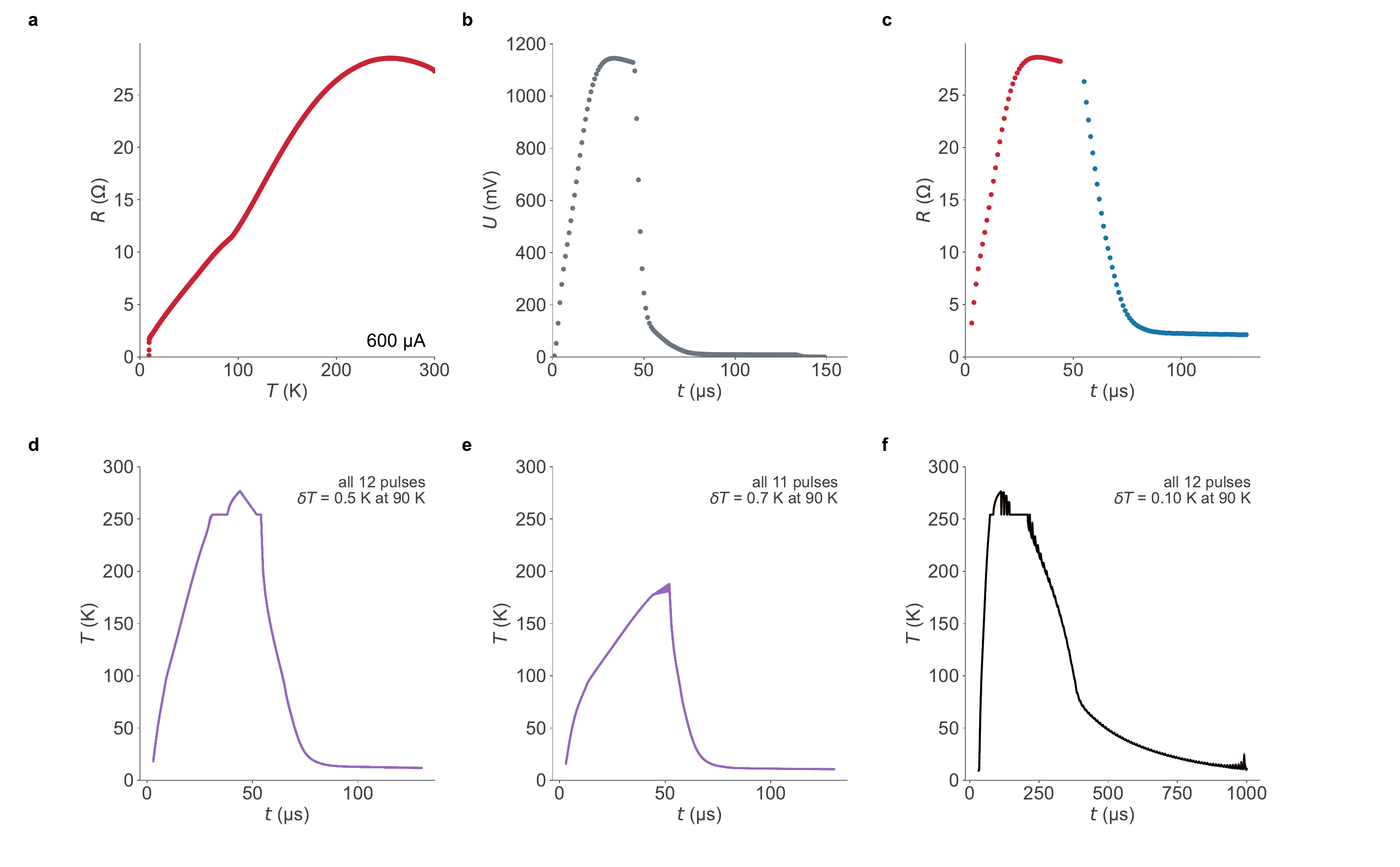}
    \caption{\textbf{Calibration and reproducibility of heat pulses} \textbf{a} Cooldown curve measured with pulsed IV at \SI{600}{\uA}. \textbf{b} Heat profile of the first \si{40}{mA} square pulse applied in the pulse train shown in the main text. After the pulse, the current is reduced to \SI{4}{mA} as fast as instrumentation allows to measure the cooldown. The maximum in $U(T)$ is evident well before the current is reduced, showing close agreement with the cooldown curve despite the thermal inhomogeneity of the sample. \textbf{c} Corresponding 4-point resistance during the pulse. \textbf{d} The heat profile of the square pulse, as well as the other 11 heat pulses shown during the pulse train, which are indistinguishable from each other. Using the timestamp closest to \SI{90}{K} we calculate the standard deviation of the instantaneous temperature between pulses at less than \SI{1}{K}. The glitch around \SI{250}{K} is due to the maximum in $R(T)$. \textbf{e} All heat profiles for the cold quench. \textbf{f} All heat pulses for the 'slow' anneal.}
\end{figure}

\clearpage
\begin{figure}
    \centering
    \includegraphics[width=\linewidth]{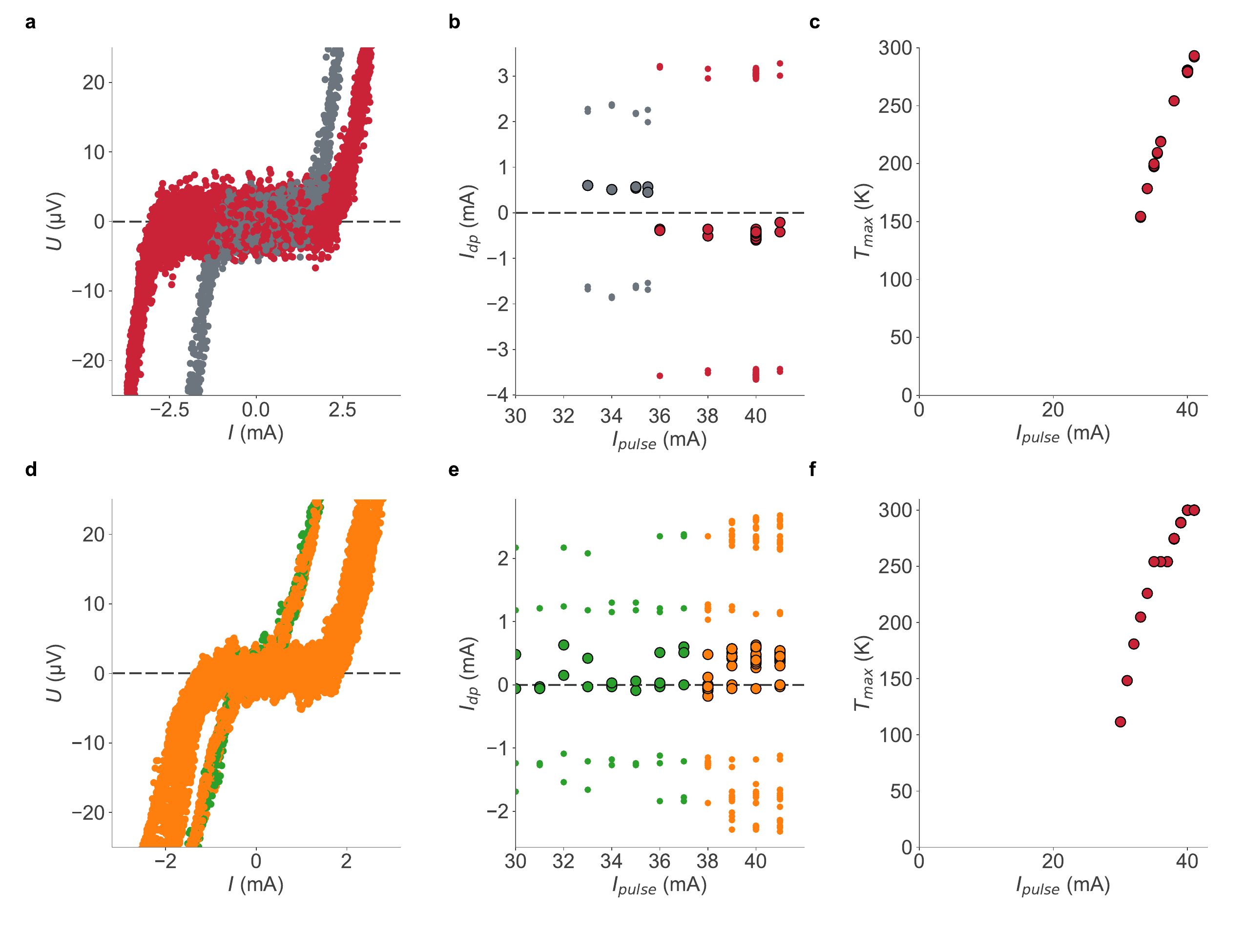}
    \caption{\textbf{Pulse design and optimization}. \textbf{a} IV curves after current pulsing using the fast-quench strategy with a square current pulse. We observe two stable configurations after many current pulses. \textbf{b} Overview of the critical current after each pulse in panel \textbf{a}. Small circles indicate critical currents for both polarities, while large marks show the diode effect. A clear cutoff around \SI{35.7}{mA} is observed. Above this current value the structure stabilizes a state with higher critical current and opposite diode sign, which we understand as the formation of a nanodomained state. \textbf{c} The peak temperature reached during each pulse. Temperature is measured by comparing the resistance during the pulse with the cooldown curve. \textbf{d} The corresponding data using a linear current ramp over \SI{1}{ms}. There are again stable states, although not deterministically reproducible such as for the fast quench strategy. \textbf{e} Overview of the critical current after each pulse in panel \textbf{d}. A current cutoff is found above which the high critical current state is more likely. \textbf{f} Peak temperatures reached during pulsing. From these data the pulse parameters shown in the main text were chosen. No further pulse shape optimization was attempted.}
\end{figure}

\clearpage
\begin{figure}
    \centering
    \includegraphics[width=\linewidth]{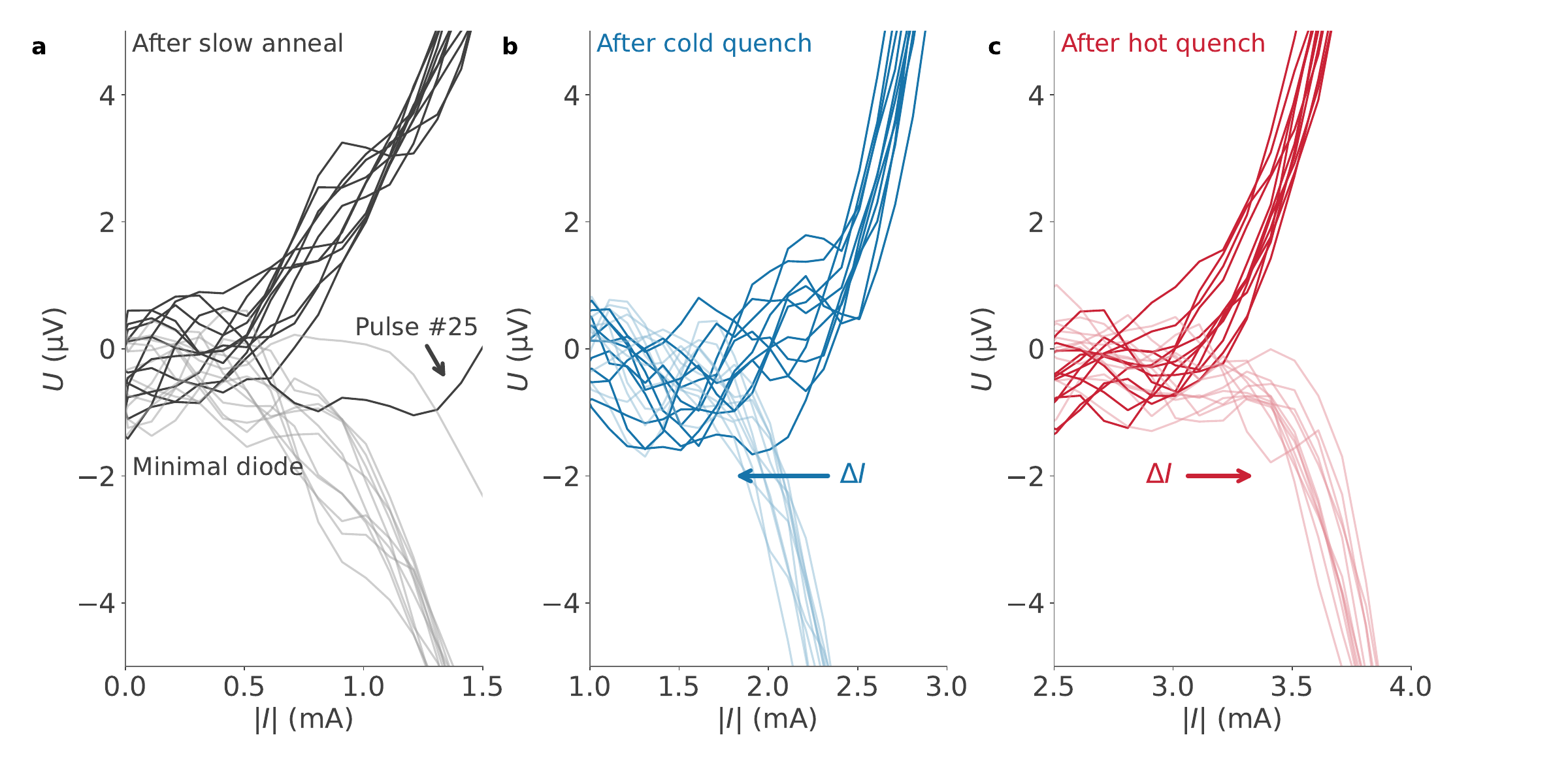}
    \caption{\textbf{Zoom of the IV curves during the pulse train} \textbf{a} Zoom-in on the IV curves obtained after slow anneals in the pulse train shown in the main text. After slow anneals the diode effect is negligible except for the anomalous pulse \#25, which appears to have failed to switch. We use $|I|$ to highlight the critical current asymmetry. \textbf{b} Same but after cold quenching. A clear SDE as well as increase in critical current is observed. \textbf{c} Same but after hot quenching. A clear SDE with opposite sign to panel \textbf{b} is observed, showing the inversion of the SDE chirality under microsecond current pulsing. All data in device~1 at \SI{0.75}{T} and \SI{6.8}{K}.}
\end{figure}

\clearpage
\begin{figure}
    \centering
    \includegraphics[width=\linewidth]{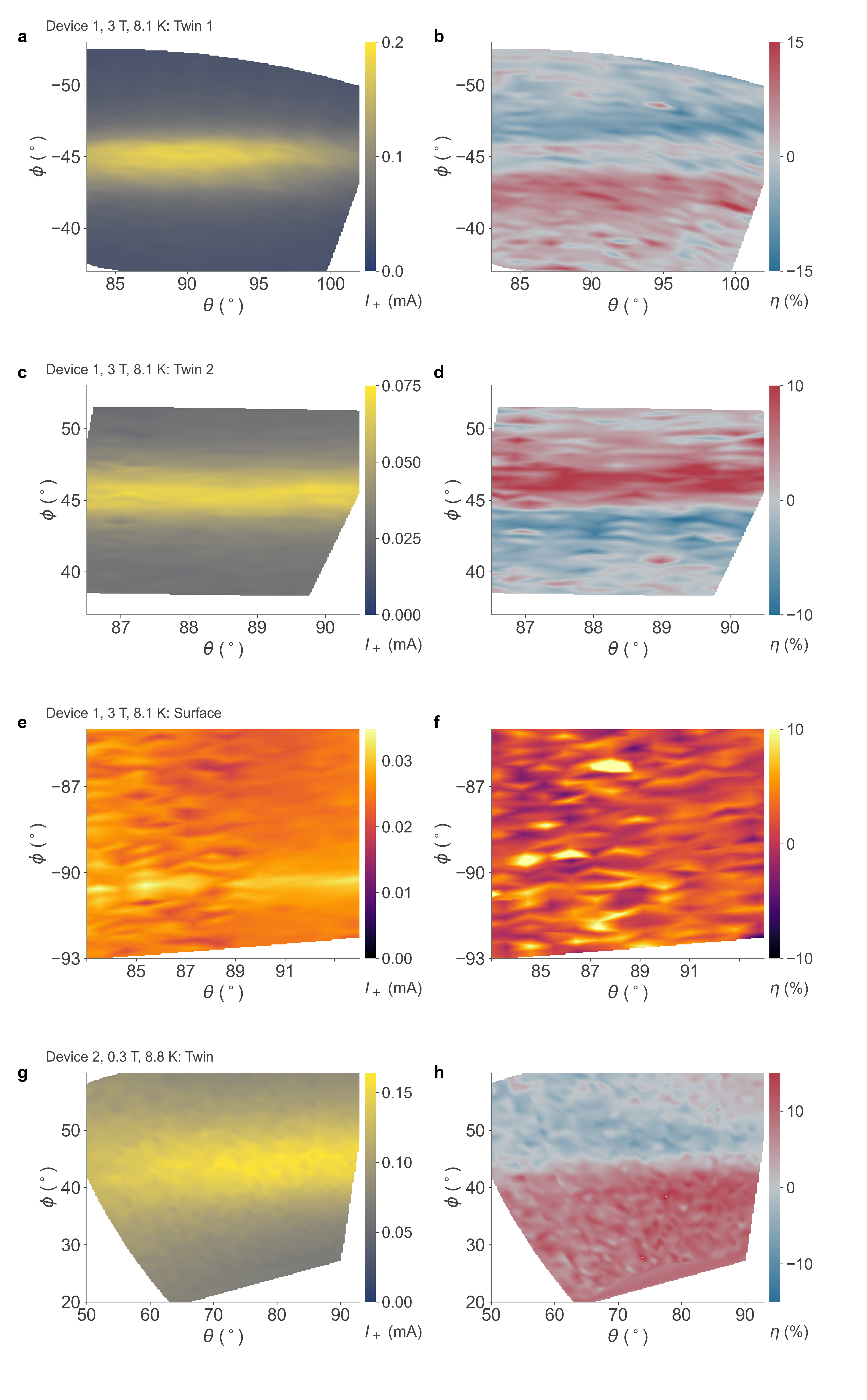}
    \caption{\textbf{Diode efficiency in the 2-axis rotator}. \textbf{a} Critical current in positive polarity for device~1 under the specified temperature and magnetic field strength. The diode is slow-cooled together with the cryostat as a whole before all maps. \textbf{b} The corresponding diode effect shows anti-symmetry in the twin domain direction as well as two additional, subtle sign changes around 44 and \ang{46} in $\phi$ direction. Meanwhile, the effect is insensitive to out-of-plane misalignment in $\theta$ direction. \textbf{cd} Similar data but for the other twin boundary direction. We show the data in the same device, angle span, temperature and applied magnetic field. This twin shows somewhat weaker overall pinning but similar diode effect size and angle range. As expected from the symmetry of the nematic domains, the sign change is in opposite direction. \textbf{ef} Again the same device under the same conditions, but now aligning the magnetic field to the surface of the microstructure. We observe a weak critical current enhancement, but no observable diode effect survives to \SI{3}{T}. \textbf{gh} Detailed map at one of the twin boundaries in device~2. We reproduce the diode effect diode effect, including its sign change in $\phi$ direction, insensitivity in $\theta$ direction and similar magnitude as device~1. While at 3~T the $\theta$ acceptance angle is $\approx$\ang{10}, at \SI{0.3}{T} it is $\approx$\ang{40}.}
\end{figure}

\clearpage
\begin{figure}
    \centering
    \includegraphics[width=\linewidth]{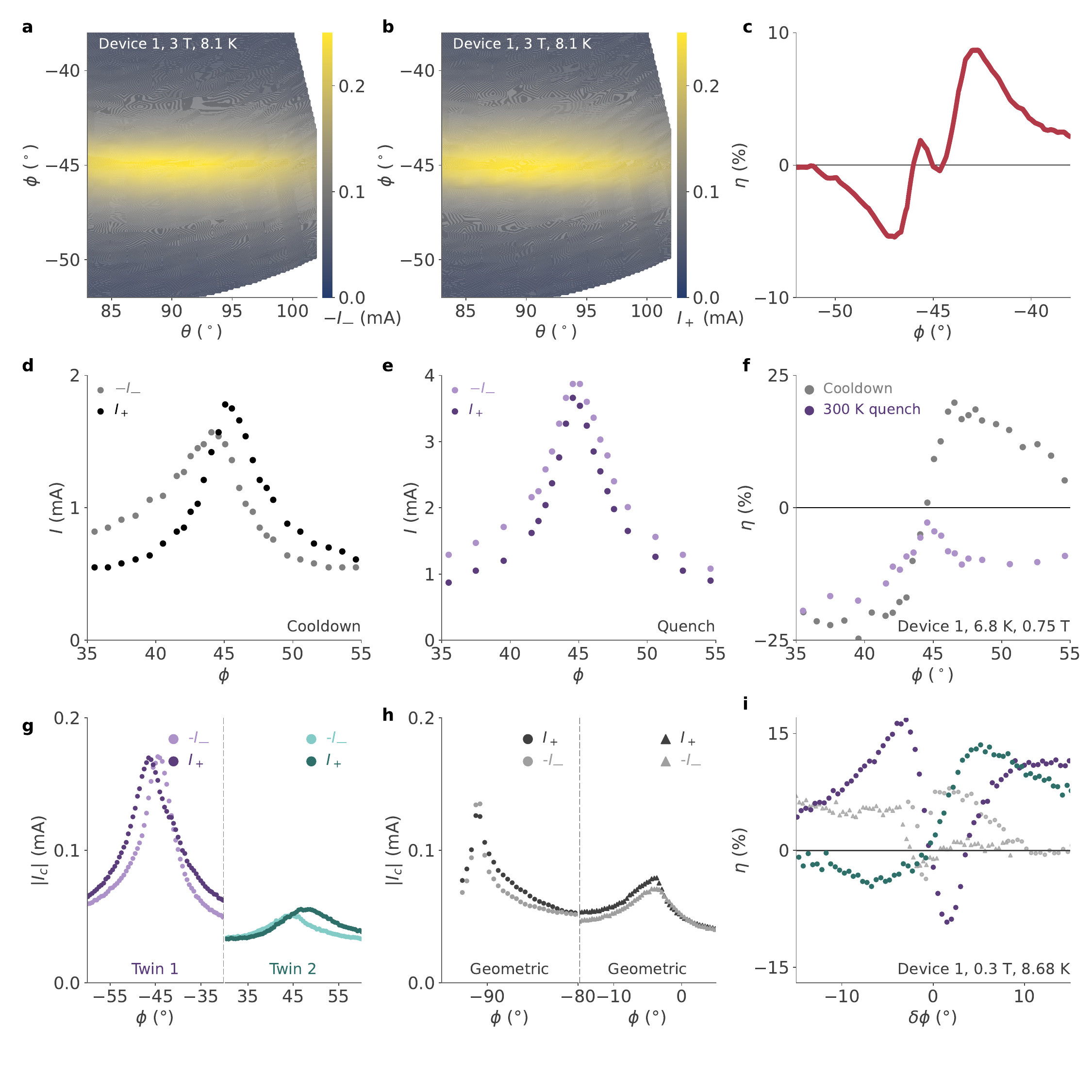}
    \caption{\textbf{Symmetries in the angle dependent SDE} \textbf{ab} The critical current maps for the diode map on the 2-axis rotator shown in the main text taken with the lock-in method. \textbf{c} Line cut at $\theta=$\ang{90}. The data show a triple sign change while maintaining odd mirror symmetry at the twin boundary direction. One explanation of this effect is different characteristic lengths for the two twin domain wall chiralities, resulting in a narrow pinning acceptance angle for leaning vortices for this type. \textbf{de} Angle dependence on cooldown and after the first \SI{40}{mA} (nearly room temperature) quench using the pulsed-IV technique. The critical current is enhanced after pulsing. \textbf{f} The mirror symmetry in the twin domain wall direction has changed. The node at the domain wall direction remains, but the diode effect obtains equal chirality on both sides with different amplitudes in this nanodomained state. \textbf{gh} Angle dependence of all diode sources after slow cooldown using the pulsed-IV technique. The two twin directions are unequal in strength due to equal number and length of domain walls between the two directions. Geometric pinning is weaker than twin pinning at \SI{0.3}{T}. \textbf{i} Corresponding diode effects. The two twin boundary directions show opposite chirality, as well as a node when the field is aligned with the domain walls. These symmetries are consistent with leaning vortices. By contrast, the geometric diode effect is maximal when the field is aligned with the surface and smaller at \SI{0.3}{T}. }
\end{figure}

\end{document}